\documentclass[useAMS,usenatbib]{mn2e}
\usepackage{graphicx}
\usepackage[usenames]{color}

\title[{\rm Kepler} photometry of RRc stars]{\textit{Kepler} photometry of RRc stars: peculiar double-mode pulsations and period doubling}
\author[P. Moskalik et al.]
{P.~Moskalik$^{1}$\thanks{E-mail: pam@camk.edu.pl},
R.~Smolec$^{1}$, K.~Kolenberg$^{2,3}$, L.~Moln\'ar$^{4,5}$, D.\,W.~Kurtz$^{6}$, R.~Szab\'o$^{4}$,
\and
J.\,M.~Benk\H{o}$^{4}$, J.\,M.~Nemec$^{7}$, M.~Chadid$^{8}$, E.~Guggenberger$^{9,10}$, C.-C.~Ngeow$^{11}$,
\and
Y.-B.~Jeon$^{12}$, G.~Kopacki$^{13}$, S.\,M.~Kanbur$^{14}$\\
\\
~$^{1}$Copernicus Astronomical Center, ul. Bartycka 18, 00-716 Warsaw, Poland\\
~$^{2}$Harvard-Smithsonian Center for Astrophysics, 60 Garden Street, Cambridge MA 02138, USA\\
~$^{3}$Instituut voor Sterrenkunde, KU Leuven, Celestijnenlaan 200D, B-3001 Heverlee, Belgium\\
~$^{4}$Konkoly Observatory, MTA CSFK, Konkoly Thege Mikl\'os \'ut 15-17, H-1121 Budapest, Hungary\\
~$^{5}$Institute of Mathematics and Physics, Savaria Campus, University of West Hungary, K\'arolyi G\'asp\'ar t\'er 4, H-9700 Szombathely, Hungary\\
~$^{6}$Jeremiah Horrocks Institute, University of Central Lancashire, Preston PR1 2HE, UK\\
~$^{7}$Department of Physics \& Astronomy, Camosun College, Victoria, BC, V8P5J2, Canada\\
~$^{8}$Universit\'e Nice Sophia-Antipolis, Observatoire de la C\^ote d'Azur, UMR 7293, Parc Valrose, 06108 Nice Cedex 02, France\\
~$^{9}$Max Planck Institut f\"ur Sonnensystemforschung, Justus-von-Liebig-Weg 3, 37077 G\"ottingen, Germany\\
$^{10}$Stellar Astrophysics Centre, Department of Physics and Astronomy, Aarhus University, Ny Munkegade 120, 8000 Aarhus C, Denmark\\
$^{11}$Graduate Institute of Astronomy, National Central University, Jhongli 32001, Taiwan\\
$^{12}$Korea Astronomy and Space Science Institute, 776, Daedeokdae-ro, Yuseong-gu, Daejeon, 305-348, Korea\\
$^{13}$Instytut Astronomiczny Uniwersytetu Wrocl{}awskiego, Kopernika 11, 51-622 Wrocl{}aw, Poland\\
$^{14}$Department of Physics, SUNY Oswego, Oswego, NY, 13126, USA\\
}

\begin{document}

\date{Accepted 2014 December 1. Received 2014 November 28; in original form 2014 October 11}

\maketitle

\begin{abstract}
We present the analysis of four first overtone RR~Lyrae stars
observed with the {\it Kepler} space telescope, based on data
obtained over nearly 2.5\,yr. All four stars are found to be
multiperiodic. The strongest secondary mode with frequency $f_2$ has
an amplitude of a few mmag, $20-45$ times lower than the main radial
mode with frequency $f_1$. The two oscillations have a period ratio
of $P_2/P_1 = 0.612-0.632$ that cannot be reproduced by any two
radial modes. Thus, the secondary mode is nonradial. Modes yielding
similar period ratios have also recently been discovered in other
variables of the RRc and RRd types. These objects form a homogenous
group and constitute a new class of multimode RR~Lyrae pulsators,
analogous to a similar class of multimode classical Cepheids in the
Magellanic Clouds. Because a secondary mode with $P_2/P_1\!\sim\!
0.61$ is found in almost every RRc and RRd star observed from space,
this form of multiperiodicity must be common. In all four {\it
Kepler} RRc stars studied, we find subharmonics of $f_2$ at $\sim\!
1/2 f_2$ and at $\sim\! 3/2 f_2$. This is a signature of period
doubling of the secondary oscillation, and is the first detection of
period doubling in RRc stars. The amplitudes and phases of $f_2$ and
its subharmonics are variable on a timescale of $10-200$\,d. The
dominant radial mode also shows variations on the same timescale,
but with much smaller amplitude. In three {\it Kepler} RRc stars we
detect additional periodicities, with amplitudes below 1\,mmag, that
must correspond to nonradial $g$-modes. Such modes never before have
been observed in RR~Lyrae variables.
\end{abstract}

\begin{keywords}
techniques: photometric -- stars: horizontal branch -- stars:
oscillations -- stars: variable: RR~Lyrae -- stars: individual:
KIC\,4064484; KIC\,5520878; KIC\,8832417; KIC\,9453114
\end{keywords}

\section{Introduction}\label{}

RR~Lyrae variables are evolved stars burning helium in their cores.
In the Hertzsprung-Russell diagram they are located at the
intersection of the horizontal branch and the classical instability
strip, in which the $\kappa$-mechanism operating in the H and He
partial ionization zones drives the pulsation. They are classified
according to their pulsation characteristics using a variant of the
initial classification by Bailey (1902). The subclass of the RRab
stars is by far the largest. These stars pulsate in the radial
fundamental mode (F), with periods of $0.3-1.0$\,d, and peak-to-peak
amplitudes in $V$ ranging from few tenths of magnitude at long
periods to more than 1\,mag at short periods. Their light curves are
asymmetric (steeper on the rising part). Their less numerous
siblings are the RRc stars, which pulsate in the first overtone
radial mode (1O) with shorter periods in the range $0.2-0.5$\,d, and
with more sinusoidal light curves with lower amplitudes of about
$0.5$\,mag in $V$, peak-to-peak. Even less numerous are the RRd
stars, which pulsate simultaneously in the radial first overtone
mode and the radial fundamental mode (F+1O).

Playing an important role in distance determination and in Galactic
structure and evolution studies, RR~Lyrae stars are among the best
studied and most observed classes of variable stars. In recent years
dozens of RR~Lyrae stars have been observed with unprecedented
precision from space by the MOST (Gruberbauer et al. 2007), CoRoT
(e.g. Szab\'o et al. 2014) and {\it Kepler} (e.g., Kolenberg et al.
2010; Benk\H{o} et al. 2010) telescopes.

Nevertheless, many intriguing puzzles surround the RR~Lyrae stars.
The most stubborn problem is the Blazhko effect, a quasi-periodic
modulation of pulsation amplitude and phase that has been known for
more than 100\,yr (Blazhko 1907). Dedicated ground-based campaigns
(Jurcsik et al. 2009) and results of {\it Kepler} observations
(Benk\H{o} et al. 2010) indicate that up to 50~per~cent of the RRab
stars show Blazhko modulation. For the RRc stars the incidence rate
is probably lower. Ground-based observations show it is below
10~per~cent (e.g. Mizerski 2003; Nagy \& Kov\'acs 2006). We lack
high-precision space observations for these stars. Only recently the
first Blazhko modulated RRc star was observed from space by {\it
Kepler} (Moln\'ar et al., in preparation). Despite many important
discoveries, including detection of period doubling (Szab\'o et al.
2010) and of excitation of additional radial modes in Blazhko
variables (Benk\H{o} et al. 2010, 2014; Moln\'ar et al. 2012) our
understanding of the Blazhko effect remains poor (for a review see
Szab\'o 2014).

Another mystery is the mode selection process in RR~Lyrae stars: we
do not know why some stars pulsate in two radial modes
simultaneously. The ability of current non-linear pulsation codes to
model this form of pulsation is still a matter of debate;  see,
e.g., Koll\'ath et al. (2002) and Smolec \& Moskalik (2008b) for
opposing views. Recent discoveries of F+2O radial double-mode
pulsations and the detection of non-radial modes in RR~Lyrae stars
(see Moskalik 2013, 2014 for reviews) make the mode-selection
problem even more topical and puzzling. Particularly interesting is
the excitation of non-radial modes, evidence of which is found in
all subclasses of RR~Lyrae variables. Their presence seems to be a
frequent phenomenon in these stars.

This paper describes new properties of first overtone RR~Lyrae
stars, some of them revealed for the first time by the
high-precision {\it Kepler} photometry. We present an in-depth study
of four RRc stars in the {\it Kepler} field: KIC\,4064484,
KIC\,5520878, KIC\,8832417 and KIC\,9453114. Partial results of our
analysis have been published in Moskalik et al. (2013). They have
also been included in a review paper of Moskalik (2014). Here we
present a full and comprehensive discussion of the results.

In Section\,\ref{Sphotom} we describe the {\it Kepler} photometry
and our methods of data reduction. Properties of the four RRc
variables are summarized in Section\,\ref{SRRc}. Our main findings
are discussed in Sections\,\ref{Sanalysis}\,--\,\ref{Sadditional},
where we present the results of frequency analyses of the stars and
describe amplitude and phase variability of the detected pulsation
modes. In Section\,\ref{Sdiscussion} we put the {\it Kepler} RRc
stars in a broader context of other recently identified multimode
RRc variables and discuss the group properties of this new type of
multimode pulsators. Our conclusions are summarized in
Section\,\ref{Ssummary}.

\section{\textit{Kepler} Photometry}\label{Sphotom}

\begin{table*}
\vskip -0.1cm
\caption{RRc stars in the {\it Kepler} field. Periods and amplitudes
         are determined from Q0+Q1 Long Cadence data.}
\label{TKeplerRRc}
\centering
\begin{tabular}{lccccccll}
\hline
KIC ID  & $\alpha$    & $\delta$       & $\langle K\!p\rangle$
                                                & Period    & $A_{\rm tot}(K\!p)$
                                                                    & [Fe/H]   & Name               & Data                   \\
        & (J\,2000)   & (J\,2000)      & [mag]  & [d]       & [mag] &          &                    &                        \\
\hline
4064484 & 19 33 45.49 & +\,39 07 14.00 & 14.641 & 0.3370019 & 0.371 & --\,1.58 &                    & Q1\,--\,Q5, Q7\,--\,Q9 \\
5520878 & 19 10 23.58 & +\,40 46 04.50 & 14.214 & 0.2691699 & 0.324 & --\,0.18 & ASAS 191024+4046.1 & Q1\,--\,Q10            \\
8832417 & 19 46 54.31 & +\,45 04 50.23 & 13.051 & 0.2485464 & 0.275 & --\,0.27 & ASAS 194654+4504.8 & Q0\,--\,Q10            \\
9453114 & 19 03 50.52 & +\,46 01 44.11 & 13.419 & 0.3660786 & 0.410 & --\,2.13 & ASAS 190350+4601.7 & Q0\,--\,Q10            \\ 
\hline
\end{tabular}
\end{table*}


The {\it Kepler} space telescope was launched on 2009 March 6 and
placed in a 372.5-d Earth trailing heliocentric orbit. The primary
purpose of the mission was to detect transits of Earth-size planets
orbiting Sun-like stars (Borucki et al. 2010). This goal was
achieved\footnote{http://kepler.nasa.gov/Mission/discoveries/candidates/}
by nearly continuous, ultraprecise photometric monitoring of nearly
200\,000 stars in the 115\,deg$^2$ field of view. A detailed
description of the mission design and its in-flight performance is
presented in Koch et al. (2010), Caldwell et al. (2010), Haas et al.
(2010), Jenkins et al. (2010a,b) and Gilliland et al. (2010). {\it
Kepler}'s primary mission ended after four years when the second
reaction wheel failed in May 2013.

\begin{figure}
\vskip 0.25cm
\centering
\resizebox{\hsize}{!}{\includegraphics{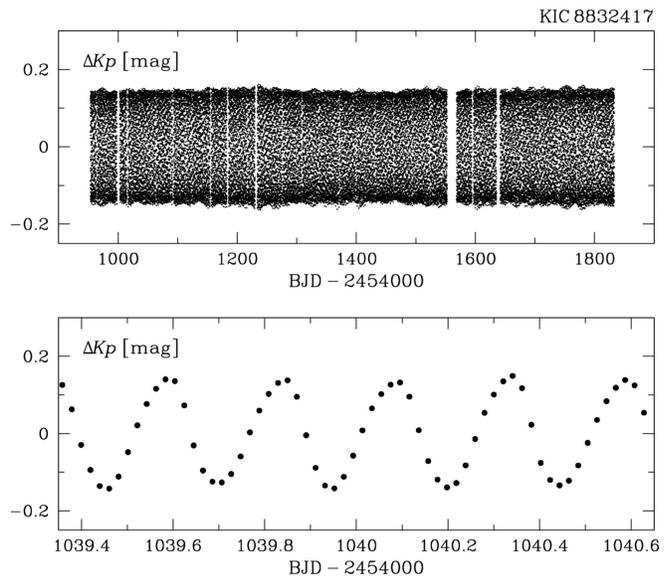}}
\caption{{\it Kepler} light curve of RRc stars KIC\,8832417 for
         Q0\,--\,Q10. Upper panel: entire light curve. Bottom panel:
         short segment of the light curve. {\it This is low
         resolution version of the figure.}}
\label{Fts}
\end{figure}

The {\it Kepler} magnitude system ($K\!p$) corresponds to a broad
spectral bandpass, from 423\,nm to 897\,nm. The time series
photometric data delivered by the {\it Kepler} telescope come in two
different formats:  Long Cadence (LC) and Short Cadence (SC), with
sampling rates of 29.43\,min and 58.86\,s, respectively. The time of
mid-point of each measurement is corrected to Barycentric Julian
Date (BJD). Four times per orbital period the spacecraft was rotated
by 90\degr\thinspace to keep optimal illumination of its solar
arrays. These rolls naturally organize the data into quarters,
denominated Q1, Q2, etc, where each quarter lasts about 3 months
(except the first and the last quarter, which are shorter). The data
are almost continuous, with only small gaps due to regular data
downlink periods and to infrequent technical problems (safe mode,
loss of pointing accuracy). The typical duty cycle of the {\it
Kepler} light curve is 92~per~cent.

\subsection{Data reduction}\label{Sreduction}

The {\it Kepler} telescope is equipped with 42 science CCDs (Koch et
al. 2010). A given star fell on four different CCDs with each
quarterly roll of the spacecraft, then returned to the original CCD.
As a result of different sensitivity levels, the measured flux jumps
from quarter to quarter. For some stars slow trends also occur
within each quarter, due to image motion, secular focus changes or
varying sensitivity of the detector (Jenkins et al. 2010a). All
these instrumental effects need to be corrected before starting the
data analysis.

Our data reduction procedure is similar to that of Nemec et al.
(2011). We use the `raw' fluxes, properly called Simple Aperture
Photometry fluxes (Jenkins et al. 2010a). The detrending is done
separately for each quarter. The flux time series is first converted
into a magnitude scale. Slow drifts in the magnitudes are then
removed by subtracting a polynomial fit. Next, the data are fitted
with the Fourier sum representing a complete frequency solution. The
residuals of this fit are inspected for any additional low-level
drifts, which are again fitted with a polynomial and subtracted from
the original magnitudes (secondary detrending). The detrended data
for each quarter are then shifted to the same average magnitude
level. As the final step, all quarters are merged, forming a
quasi-continuous $K\!p$ magnitude light curve of a star. An example
of a reduced light curve is displayed in Fig.\,\ref{Fts}.

\section{RRc stars in the \textit{Kepler} field}\label{SRRc}

More than 50 RR~Lyrae variables are currently identified in the {\it
Kepler}\, field (Kolenberg et al. 2014). At the time this study was
initiated, only four of them were classified as RRc pulsators. Basic
characteristics of these four stars stars are given in
Table\,\ref{TKeplerRRc}. The first four columns of the table list
the star numbers, equatorial sky coordinates ($\alpha,\delta$) and
mean $K\!p$ magnitude, all taken from the {\it Kepler} Input
Catalogue (KIC, Brown et al. 2011). Columns 5 and 6 contain the
pulsation periods and the total (peak-to-peak) amplitudes. All four
RRc stars turned out to be multiperiodic (Section\,\ref{Sanalysis}),
but they are all strongly dominated by a single radial mode. The
periods and amplitudes given in the table correspond to this
dominant mode of pulsation. They are determined from the {\it
Kepler} light curves. Column 7 contains spectroscopic metal
abundances, [Fe/H], from Nemec et al. (2013). Two stars in our RRc
sample are metal-rich (KIC\,5520878 and KIC\,8832417), the other two
are metal-poor. Column 8 indicates other identifications of the
stars. KIC\,9453114 is also known as ROTSE1\,J190350.47+460144.8.
None of the {\it Kepler} RRc variables has a GCVS name yet.

The last column of Table\,\ref{TKeplerRRc} lists {\it Kepler}
observing runs analysed in this paper.  Here we use only the Long
Cadence photometry, collected in quarters Q1 to Q10. For
KIC\,8832417 and KIC\,9453114 it is supplemented by 10\,d of
commissioning data (Q0). Due to the loss of CCD module no.~3, no
photometry was obtained for KIC\,4064484 in Q6 and Q10. The total
time base of the data ranges from 774\,d for KIC\,4064484 to 880\,d
for KIC\,8832417 and KIC\,9453114.

\begin{figure}
\vskip 0.25cm
\centering
\resizebox{\hsize}{!}{\includegraphics{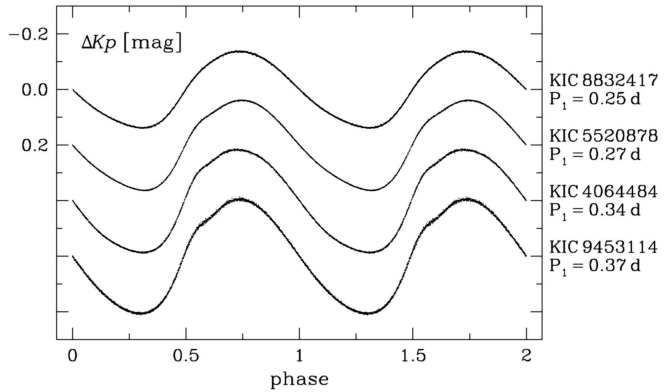}}
\caption{Phased light curves of the dominant radial mode of
         {\it Kepler} RRc stars. {\it This is low resolution
         version of the figure.}}
\label{FphasedLC}
\end{figure}
\begin{table}
\vskip -0.1cm
\caption{Light curve shape parameters for the dominant radial mode
         of {\it Kepler} RRc stars. Values are determined from Q0+Q1
         data.}
\label{TFourierRRc}
\centering
\begin{tabular}{lcccccc}
\hline
KIC     & $\log\! P_1$ & $A_1$  & $R_{21}$ & $\phi_{21}$ & $M\!- m$  \\
        & [d]          & [mmag] &          & [rad]       &           \\
\hline
8832417 & --\,0.605    & 138.40 & 0.1015   & 4.661       & 0.425     \\
5520878 & --\,0.570    & 162.88 & 0.1078   & 4.704       & 0.427     \\
4064484 & --\,0.472    & 190.50 & 0.1109   & 4.813       & 0.408     \\
9453114 & --\,0.436    & 206.38 & 0.0982   & 4.828       & 0.430     \\
\hline
\end{tabular}
\end{table}


Fig.\,\ref{FphasedLC} displays phased light curves of the dominant
mode of {\it Kepler} RRc stars. The plot is constructed with Q0+Q1
data, which were prewhitened by all frequencies other than the
dominant one ($f_1\!=\!1/P_1$) and its harmonics
(Section\,\ref{Sanalysis}). The light curves are typical for radial
first overtone pulsators, with round tops, long rise times and low
amplitudes. The change of slope which occurs shortly before the
brightness maximum is also characteristic for the RRc variables
(e.g. Lub 1977; Olech et al. 2001).

To describe light curve shape in a quantitative way we resort to a
Fourier decomposition (Simon \& Lee 1981). We fit the light curves of
Fig.\,\ref{FphasedLC} with the Fourier sum of the form

\begin{equation}
K\!p(t) = A_0 + \sum_{k} A_{k} \cos (2\pi kf_1 t + \phi_{k}).
\label{EFoursum}
\end{equation}

\noindent and then compute the usual Fourier parameters: the
amplitude ratio $R_{21}\! =\! A_2/A_1$ and the phase difference
$\phi_{21}\! =\!\phi_2-2\phi_1$. They are listed in
Table\,\ref{TFourierRRc}, together with the semi-amplitude of the
dominant frequency, $A_1$ ($\approx\!\! A_{\rm tot}/2$). The last
column of the table gives another light curve parameter: the
interval from minimum to maximum, expressed as a fraction of the
pulsation period. This quantity measures asymmetry of the light
curve and is traditionally called $M-m$ (Payne-Gaposchkin \&
Gaposchkin 1966) or a risetime parameter (e.g. Nemec et al. 2011).
For all four variables in Table\,\ref{TFourierRRc} this parameter is
above 0.4, which agrees with their classification as RRc stars
(Tsesevich 1975).

\begin{figure}
\vskip 0.20cm
\centering
\resizebox{\hsize}{!}{\includegraphics{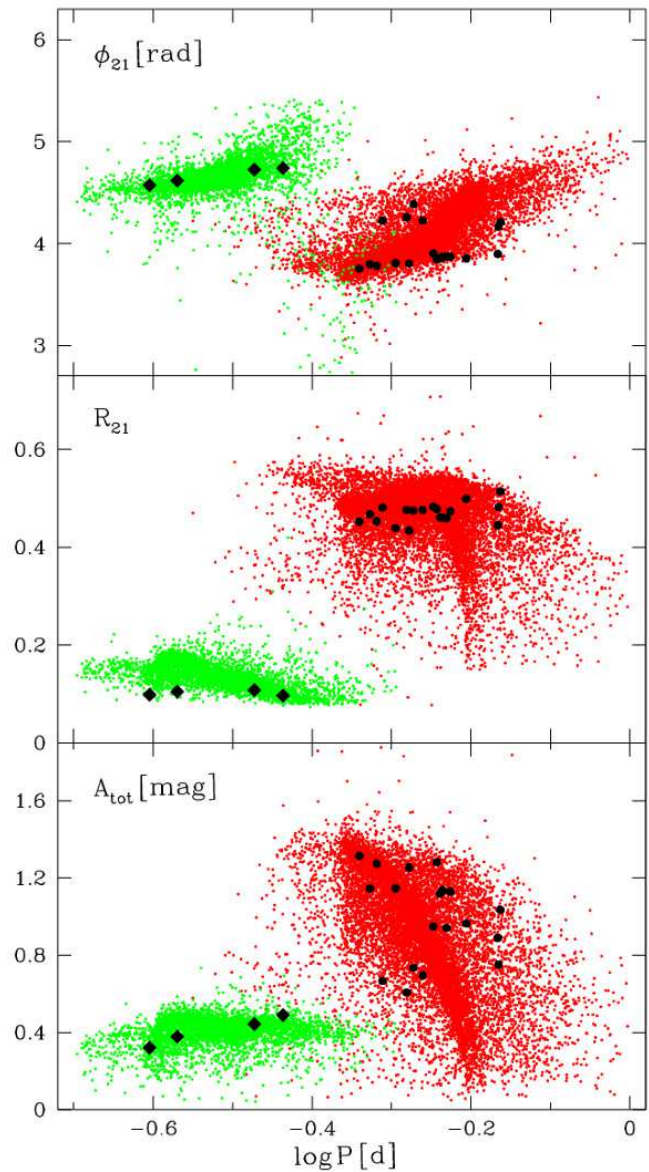}}
\caption{Fourier parameters ($V$-band) of {\it Kepler} RRc stars (black
         diamonds) and non-Blazhko RRab stars (black circles). The
         Galactic Bulge RR~Lyrae stars (Soszy\'nski et al. 2011a)
         are displayed for comparison, with the first overtone (RRc)
         and the fundamental mode (RRab) variables plotted as green
         and red dots, respectively. {\it This is low resolution version
         of the figure.}}
\label{FFourier}
\end{figure}

In Fig.\,\ref{FFourier} we compare the Fourier parameters and the
peak-to-peak amplitudes $A_{\rm tot}$ of the {\it Kepler} RRc stars
with those of RR~Lyrae stars in the Galactic Bulge (Soszy\'nski et
al. 2011a). {\it Kepler} non-Blazhko RRab stars (Nemec et al. 2011)
are also displayed. All parameters are converted to a common
photometric band, namely to Johnson $V$. For the Bulge variables,
$R_{21}$ and $\phi_{21}$ are transformed from $I$ to $V$ with the
formula of Morgan, Simet \& Bargenquast (1998). For the amplitude,
we use equation $A_{\rm tot}(V)  = 1.62 A_{\rm tot}(I)$, derived by
us from photometry of RR~Lyrae stars in M68 (Walker 1994). The
amplitude transformation is the same for RRab and for RRc stars. In
case of the {\it Kepler} variables, $\phi_{21}$ is transformed from
$K\!p$ to $V$ according to Eq.\,2 of Nemec et al. (2011), but for
$A_{\rm tot}$ and $R_{21}$ we used proportional scaling, which in
our opinion is more appropriate. Using the same data as Nemec et al.
(2011), we derive $A_{\rm tot}(V)  = 1.16 A_{\rm tot}(K\!p)$ and
$R_{21}(V)\! =\! 0.975 R_{21}(K\!p)$.

All $K\!p \rightarrow V$ transformation formulae are calibrated with
RRab stars, thus applying them to the RRc stars can yield only
approximate results. This approximation is still sufficiently
accurate, particularly for $A_{\rm tot}$ (where scaling is
mode-independent) and for $\phi_{21}$ (where colour-to-colour
corrections are always small). We expect the approximation to be
least accurate in the case of $R_{21}$, for which colour-to-colour
scaling is somewhat dependent on the pulsation mode (Morgan et al.
1998).

Fig.\,\ref{FFourier} shows that the Fourier parameters of variables
listed in Table\,\ref{TKeplerRRc} (black diamonds) are very
different from those of the {\it Kepler} RRab stars. At the same
time, they match typical values for the first overtone RR~Lyrae
stars  well. The Figure proves that the variables of
Table\,\ref{TKeplerRRc} belong to the population of RRc stars.

\section{Data analysis and results }\label{Sanalysis}

\begin{figure*}
\vskip 0.2cm
\centering
\resizebox{\hsize}{!}{\includegraphics{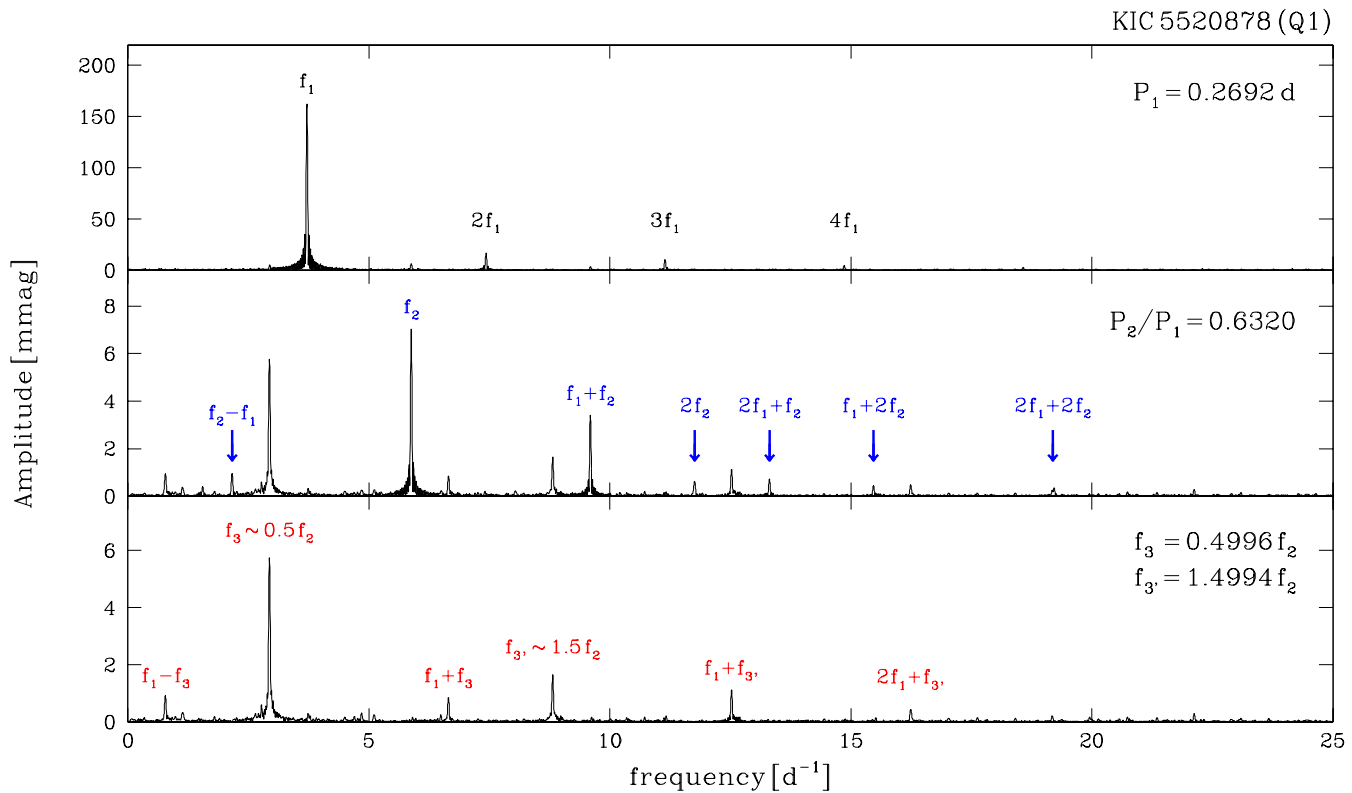}}
\caption{Prewhitening sequence for KIC\,5520878. The upper panel
         displays the Fourier transform of the original $K\!p$ magnitude
         light curve. The middle and bottom panels show FTs after
         consecutive prewhitening steps (see text). Only Q1 data are
         analyzed here.}
\label{FKIC55prew}
\end{figure*}

\begin{table*}
\vskip -0.1cm
\caption{Main frequency components identified in Q1 lightcurve of
         KIC\,5520878.}
\label{TKIC55freq}
\centering
\begin{tabular}{ccccccccccc}
\hline
frequency & $f$\,[d$^{-1}$]   & $A$\,[mmag]  &~~~~~& frequency     & $f$\,[d$^{-1}$] & $A$\,[mmag]  &~~~~~& frequency        & $f$\,[d$^{-1}$] & $A$\,[mmag] \\
\hline
~$\,f_1$  & $\,$~3.715126     &      162.88  & & ~$\,f_1-f_2$~     & $\,$~2.16017    &     0.95     & & ~~~~~~~~$0.5f_2$ &    2.93694      &     5.74    \\
  $2f_1$  & $\,$~7.430253     &  $\,$~17.55  & &   $2f_1-f_2$~     & $\,$~1.55180    &     0.41     & & ~~~~~~~~$1.5f_2$ &    8.81428      &     1.65    \\
  $3f_1$  &     11.145379     &  $\,$~11.06  & & ~$\,f_1+f_2$~     & $\,$~9.59625    &     3.37     & &  ~$\,f_1-0.5f_2$ &    0.77949      &     0.95    \\
  $4f_1$  &     14.860506     &     ~~~4.77  & &   $2f_1+f_2$~     &     13.31087    &     0.76     & &    $2f_1-0.5f_2$ &    4.48680      & $\;$0.18:   \\
  $5f_1$  &     18.575632     &     ~~~2.61  & &   $3f_1+f_2$~     &     17.02511    & $\;$0.16:    & &  ~$\,f_1+0.5f_2$ &    6.64948      &     0.87    \\
  $6f_1$  &     22.290759     &     ~~~1.39  & &   $4f_1+f_2$~     &     20.73973    & $\;$0.16:    & &  ~$\,f_1-1.5f_2$ &    5.10593      &     0.26    \\
  $7f_1$  &     26.005885     &     ~~~0.78  & & ~~$\,\,f_1-2f_2$~ & $\,$~8.03827    &     0.26     & &  ~$\,f_1+1.5f_2$ &   12.52531~$\,$ &     1.13    \\
  $8f_1$  &     29.721011     &     ~~~0.37  & & ~~$\,\,f_1+2f_2$~ &     15.46826    &     0.44     & &    $2f_1+1.5f_2$ &   16.24251~$\,$ &     0.44    \\
~$\,f_2$  &      5.87858      &     ~~~7.04  & &   ~$\,2f_1+2f_2$~ &     19.18085    & $\;$0.18:    & &    $3f_1+1.5f_2$ &   19.95592~$\,$ & $\;$0.16:   \\
  $2f_2$  &     11.75877~$\,$ &     ~~~0.65  & & ~~$\,\,f_1+3f_2$~ &     21.35066    & $\;$0.14:    & &  ~$\,f_1+2.5f_2$ &   18.41040~$\,$ & $\;$0.14:   \\
  $3f_2$  &     17.62523~$\,$ & $\;$~~~0.13: & &                   &                 &              & &    $2f_1+2.5f_2$ &   22.12452~$\,$ &     0.29    \\
\hline
\end{tabular}
\end{table*}

We analyzed the pulsations of {\it Kepler} RRc stars with a standard
Fourier transform (FT) combined with the multifrequency
least-squares fits and consecutive prewhitening. We used well-tested
software written by Z.~Ko{\l}aczkowski (see Moskalik \&
Ko{\l}aczkowski 2009). The light curve was first fitted with a
Fourier series representing variations with the dominant frequency,
$f_1$ (Eq.\,\ref{EFoursum}). After subtracting the fitted function
(prewhitening), the residuals were searched for secondary periods.
This was done with the FT, computed over the range from 0 to
24.5\,d$^{-1}$, which is the Nyquist frequency of the Long Cadence
data. Newly identified frequencies were included into the Fourier
series, which was fitted to the light curve again. The residuals of
the fit were searched for more frequencies. The whole procedure was
repeated until no new periodicities were found. In each step, all
frequencies were optimized by the least-squares routine. The final
fit yields frequencies, amplitudes and phases of all identified
harmonic components.

All four {\it Kepler} RRc variables turned out to be multiperiodic.
Very early in the analysis, we realized that the amplitudes and
phases of the detected frequencies are not constant. Therefore, we
first discuss only a short segment of available data, namely Q0+Q1.
Our goal here is to establish the frequency content of the RRc light
curves. We defer the analysis of the entire Q0\,--\,Q10 datasets and
discussion of temporal behaviour of the modes to
Section\,\ref{Svar}. In the following subsections, we present the
results for each of the investigated RRc stars individually. We
discuss the stars in order of decreasing amplitude stability, which
nearly coincides with the order of increasing pulsation period.

\subsection{KIC\,5520878}\label{SKIC55}

\begin{figure*}
\vskip 3.25truecm
\centering
\resizebox{\hsize}{!}{\includegraphics{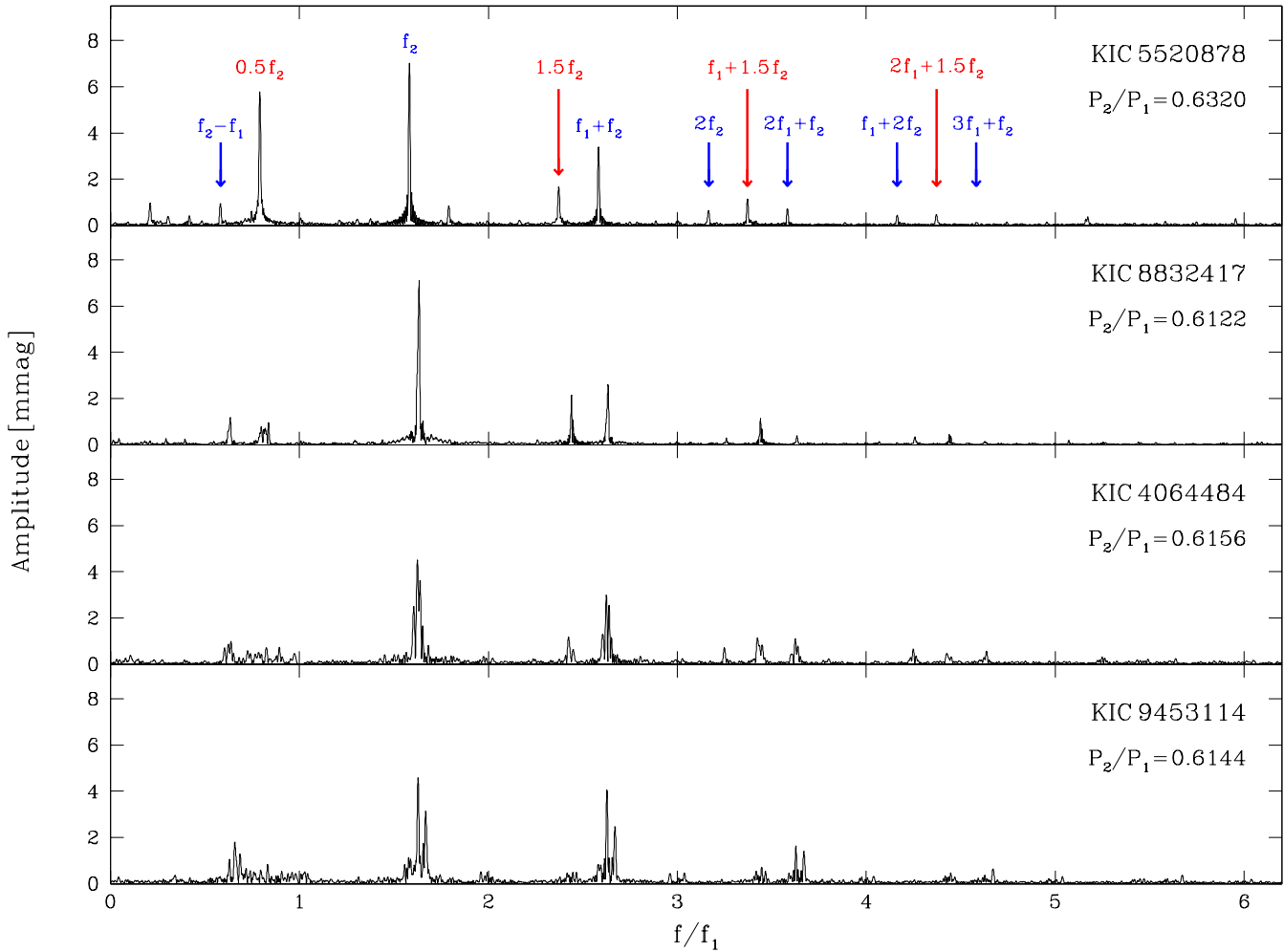}}
\caption{Frequency spectra of the four {\it Kepler} RRc stars (Q0+Q1
         data only) after pre-whitening by $f_1$ and its harmonics.
         FT of KIC\,5520878 (upper panel) is the same as plotted in
         the middle panel of Fig.\,\ref{FKIC55prew}.}
\label{FRRcprew}
\end{figure*}

Fig.\,\ref{FKIC55prew} shows the prewhitening sequence for
KIC\,5520878. The FT of its light curve (upper panel) is strongly
dominated by the radial mode with frequency $f_1\! =\!
3.715126$\,d$^{-1}$ and amplitude $A_1\! =\! 162.88$\,mmag. After
subtracting $f_1$ and its harmonics from the data, the FT of the
residuals (middle panel) shows the strongest peak at $f_2\! =\!
5.87858$\,d$^{-1}$. This secondary mode has an amplitude of only
$A_2\! =\! 7.04$\,mmag, 23 times lower than $A_1$. Its harmonic and
several combinations with $f_1$ are also clearly visible. The period
ratio of the two modes is $P_2/P_1\! =\! 0.6320$. After prewhitening
the light curve with both $f_1$, $f_2$ and their harmonics and
combinations (bottom panel), the strongest remaining signal appears
at $f_3\! =\! 2.93694$\,d$^{-1}$, i.e. at $\sim$1/2$f_2$. Thus,
$f_3$ is not an independent frequency, it is a {\it subharmonic} of
$f_2$. A second subharmonic at $f_{3'}\! =\! 8.81428$\,d$^{-1}\!
\sim\! 3/2f_2$ is also present. The remaining peaks in the FT
correspond to linear combinations of $f_1$ with $f_3$ and $f_{3'}$.
In Table\,\ref{TKIC55freq} we list all harmonics, subharmonics and
combination frequencies identified in Q1 light curve of
KIC\,5520878.

\subsection{KIC\,8832417}\label{SKIC88}

Pulsations of this star are dominated by the radial mode with
frequency $f_1\! =\! 4.02339$\,d$^{-1}$. After prewhitening the data
with $f_1$ and its harmonics (Fig.\,\ref{FRRcprew}, second panel) we
identify a secondary mode at $f_2 = 6.57203$\,d$^{-1}$. The
resulting period ratio is $P_2/P_1 = 0.6122$. The subharmonic of the
secondary frequency at $\sim\! 3/2 f_2$ is clearly visible. The
subharmonic at $\sim$1/2$f_2$ is present as well, but its appearance
is different. The peak is very broad and almost split into three
close components. Such a pattern implies that we look either at a
group of frequencies that are too close to be resolved, or at a
single frequency with unstable amplitude and/or phase. The central
component of this subharmonic power is located at $f_{3} =
3.29848$\,d$^{-1} = 0.5019 f_2$. The remaining peaks in the
prewhitened FT correspond to linear combinations of $f_1$ with $f_2$
and $3/2 f_2$.

\subsection{KIC\,4064484}\label{SKIC40}

Pulsations of KIC\,4064484 are dominated by the radial mode with
frequency $f_1\! =\! 2.96734$\,d$^{-1}$. After prewhitening the data
with $f_1$ and its harmonics (Fig.\,\ref{FRRcprew}, third panel) we
find a secondary mode at $f_2 = 4.82044$\,d$^{-1}$, yielding a
period ratio of $P_2/P_1\! =\! 0.6156$. The peak corresponding to
$f_2$ is broadened and starts to split into several unresolved
components. The same is true for the combination peaks at $f_2 -
f_1$, $f_1$\,+\,$f_2$, $2 f_1$\,+\,$f_2$, etc. As in the other two
RRc stars, we find a subharmonic of the secondary frequency at
$\sim\! 3/2 f_2$. This peak is also broadened and split, and so are
the corresponding combination peaks. Unlike the previous two RRc
stars, in KIC\,4064484 we do not detect a clear subharmonic at
$\sim\! 1/2 f_2$. There is an excess of power in the vicinity of
this frequency, but it forms a dense forest of peaks, none of which
stand out.

\subsection{KIC\,9453114}\label{SKIC94}

\begin{table*}
\vskip -0.1cm
\caption{Primary and secondary periodicities in {\it Kepler} RRc
         stars, determined from Q0+Q1 data.}
\label{TRRcfreq}
\centering
\begin{tabular}{lccccccccc}
\hline
KIC     & $P_1$     & $A_1$  &~& $P_2$   & $A_2$  &  $P_{3,3'}$   &  $A_{3,3'}$ & $P_2/P_1$ &  $f_{3,3'}/f_2$  \\
        & [d]       & [mmag] & & [d]     & [mmag] &  [d]          &  [mmag]     &           &                  \\
\hline
8832417 & 0.2485464 & 138.40 & & 0.15216 & 7.11   &  0.30317      &  0.69       & 0.61218   &  0.5019          \\ 
        &           &        & &         &        &  0.10186      &  2.14       &           &  1.4937          \\ 
5520878 & 0.2691699 & 162.88 & & 0.17011 & 7.04   &  0.34049      &  5.74       & 0.63197   &  0.4996          \\ 
        &           &        & &         &        &  0.11345      &  1.65       &           &  1.4994          \\ 
4064484 & 0.3370019 & 190.50 & & 0.20745 & 4.55   &  0.13901      &  1.21       & 0.61558   &  1.4923          \\ 
9453114 & 0.3660809 & 206.64 & & 0.22490 & 4.55   & (0.43987)     & (0.81)      & 0.61435   & (0.5113)         \\ 
        &           &        & &         &        &  0.14959      &  0.46       &           &  1.5034          \\ 
\hline
\end{tabular}
\end{table*}



The dominant radial mode of KIC\,9453114 has a frequency of $f_1 =
2.73164$\,d$^{-1}$. After prewhitening the light curve with $f_1$ and
its harmonics (Fig.\,\ref{FRRcprew}, bottom panel) we find a
secondary mode at $f_2\! =\! 4.44643$\,d$^{-1}$, yielding a period
ratio of $P_2/P_1 = 0.6143$. A second peak of somewhat lower
amplitude is present next to it, at $f_{2'}\! = 4.55830$\,d$^{-1}$.
Another much weaker peak can be identified at the same distance from
$f_2$, but on its opposite side. Thus, the secondary frequency in
KIC\,9453114 is split into three {\it resolved components}, which
form an approximately equidistant triplet. The same pattern is also
seen at the combination frequencies $f_2 - f_1$, $f_1$\,+\,$f_2$, $2
f_1$\,+\,$f_2$, etc. The $f_2$ triplet can be interpreted in two
physically different ways: either as a multiplet of nonradial modes
($\ell \ge 1$) split by rotation or as a single mode undergoing a
periodic or quasi-periodic modulation. A close inspection of the
triplet shows that all three components are incoherent. Indeed, an
attempt to prewhiten them with three sine waves of constant
amplitudes and phases proves unsuccessful, leaving significant
residual power. This casts doubt on the nonradial multiplet
interpretation. We will return to this point in
Section\,\ref{Sf2freq}.

Although very weak, the subharmonics of $f_2$ are also present in
KIC\,9453114. The signal corresponding to $\sim\! 3/2 f_2$ is
detected at $f_{3'} = 6.68487$\,d$^{-1}$. Its combination
frequencies at $\sim\!\! f_1\! +\! f_{3'}$ and $\sim\!\! 2 f_1\! +\!
f_{3'}$ can be identified, too. All these peaks are split into
resolved triplets. A second subharmonic ($\sim$$1/2 f_2$) is
detected as well, as a marginally significant single peak at $f_3 =
2.27340$\,d$^{-1}$.

\subsection{Similarity of \textit{Kepler} RRc stars}\label{Ssimilar}

The results of the frequency analyses of {\it Kepler} RRc variables
are summarized in Table\,\ref{TRRcfreq}. In Fig.\,\ref{FRRcprew} we
display the Fourier transforms of their Q0+Q1 light curves,
prewhitened of the dominant radial mode. For better comparison, the
FTs are plotted vs. normalized frequency, $f/f_1$.

All four RRc variables are multiperiodic. Fig.\,\ref{FRRcprew}
shows that their frequency spectra are remarkably similar and
display peaks at the same places. In each star we detect a secondary
mode, which appears at $f_2/f_1 = 1.58-1.63$ or $P_2/P_1 =
0.612-0.632$. This is always the highest secondary peak, yet its
amplitude is only a few mmag and is $20-45$ times lower than the
amplitude of the radial mode. Without the benefit of the high-precision
{\it Kepler} photometry, such a weak signal is very difficult to detect.

In each variable we identify at least one subharmonic of the
secondary frequency. The signal at $\sim\! 3/2 f_2$ is detected in
all {\it Kepler} RRc stars. The second subharmonic at $\sim$$1/2
f_2$ is visible in three of the stars, although it is prominent only
in KIC\,5520878. The presence of subharmonics, i.e., frequencies of
the form $(n+1/2)f$, is a characteristic signature of a period
doubling (Berg\'e et al. 1986; see also Fig.\,3 of Smolec \&
Moskalik 2012). Thus, our finding constitutes the first detection of
the period doubling phenomenon in the RRc variables (see also
Moskalik et al. 2013; Moskalik 2014). These stars are thus the
fourth class of pulsators in which period doubling has been
discovered, following the RV~Tauri stars (known for decades), and
the Blazhko RRab stars (Kolenberg et al. 2010, Szab\'o et al. 2010)
and the BL~Herculis stars (Smolec et al. 2012), only identified
recently.

We note, that the subharmonics listed in Table\,\ref{TRRcfreq} are
never located at precisely $1/2 f_2$ and $3/2 f_2$. The deviations
from the exact half-integer frequency ratios are very small, almost
never exceeding 0.5 per cent, but they are statistically
significant. We recall, that similar deviations are also observed
for period doubling subharmonics in the Blazhko RRab stars (Szab\'o
et al. 2010, 2014; Kolenberg et al. 2011; Guggenberger et al. 2012).
This behaviour has been traced to the nonstationary character of the
subharmonics, which causes their instantaneous frequencies to
fluctuate around the expected values (Szab\'o et al. 2010). The same
reasoning applies also to the RRc stars. As we will discuss in the
next section, the subharmonics detected in these variables are
nonstationary as well.

Finally, we note that a secondary mode with $P_2/P_1 \sim 0.61$ and
its subharmonics are detected in {\it every} RRc star observed by
{\it Kepler}. This suggests that excitation of this mode and the
concomitant period doubling is not an exception, but is a common
property of the RRc variables. We return to this point in
Section\,\ref{Sdiscussion}.

\section{Variability of amplitudes and phases}\label{Svar}

\subsection{$\bmath{f_2}$ and its subharmonics}\label{}

We start the discussion with the secondary mode, $f_2$, for which
evidence of instability is most noticeable. Fig.\,\ref{FRRcprew}
presents RRc frequency spectra computed for month-long subsets of
available data. Already for this short timebase, the Fourier
peaks corresponding to $f_2$ and its subharmonics are broadened or
even split. This indicates that amplitudes and/or phases of these
frequency components are variable. To examine this variability in
detail, in this Section we analyze for each star the entire
light curve, covering quarters Q0\,-- Q10.

\subsubsection{Time domain}\label{}

\begin{figure}
\vskip 0.3truecm
\centering
\resizebox{\hsize}{!}{\includegraphics{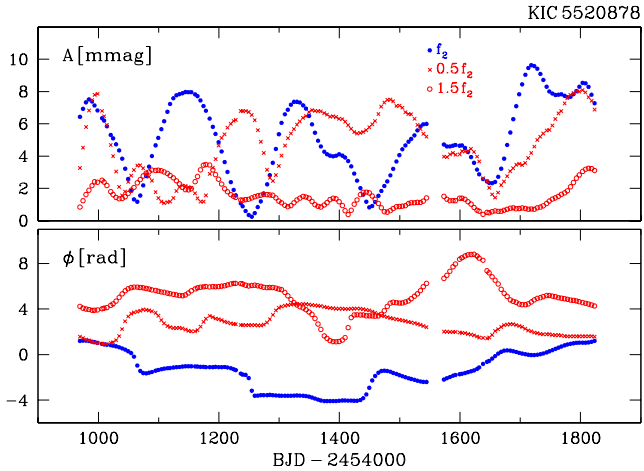}}
\caption{Variability of the secondary mode and its subharmonics in
         KIC\,5520878. Upper panel: amplitude variations. Bottom
         panel: phase variations.}
\label{FKIC55amp}
\end{figure}

We examine the temporal behaviour of modes by applying a
time-dependent Fourier analysis (Kov\'acs, Buchler \& Davis 1987).
To this end, we subdivide the light curve into short overlapping
segments with duration $\Delta t$, and then fit a Fourier series
consisting of all significant frequency terms to each segment
separately. All frequencies are kept fixed. The choice of $\Delta t$
is somewhat arbitrary and depends on how fast the amplitudes and
phases change. With this procedure, we can follow the temporal
evolution of all frequency components present in the data.

In Fig.\,\ref{FKIC55amp} we present results of such an analysis for
KIC\,5520878. For this star we adopted $\Delta t=\!10$\,d. The plot
displays amplitudes and phases of the secondary mode, $f_2$, and of
its two subharmonics,  $1/2f_2$ and $3/2f_2$. The amplitude of the
secondary mode varies in a rather irregular fashion, with a
timescale of $\sim$\,200\,d. The range of these variations is
extremely large: from almost zero to 9.6\,mmag. In other words, the
amplitude of $f_2$ fluctuates by nearly 100~per~cent! Both
subharmonics of $f_2$ display large, irregular changes as well.
Interestingly, although occurring with approximately the same
timescale, they do not seem to be correlated with variations of the
parent mode -- the maximum (minimum) amplitudes of the subharmonics
in some instances coincide with maximum (minimum) amplitude of
$f_2$, but in other instances they do not. The amplitude variability
of the secondary mode and its subharmonic is accompanied by
irregular variability of their phases (Fig.\,\ref{FKIC55amp}, bottom
panel).

\begin{figure}
\vskip 1.8truecm
\centering
\resizebox{\hsize}{!}{\includegraphics{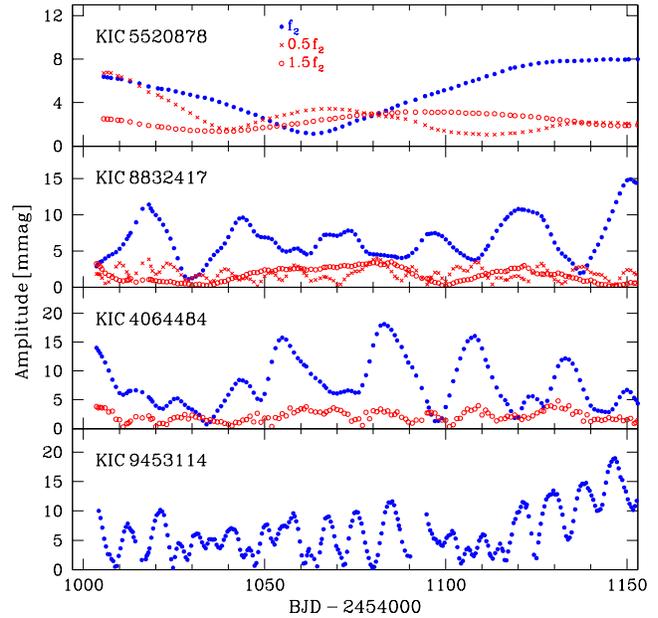}}
\caption{Amplitude variations of the secondary mode and its
         subharmonics in {\it Kepler} RRc stars. For better
         visibility, amplitudes of subharmonics in KIC\,8832417
         (KIC\,4064484) are multiplied by 1.5 (by 2.0). Subharmonics
         of KIC\,9453114 are too weak to be securely extracted from
         short segments of data. Only part of total data is displayed.}
\label{FRRcamp}
\end{figure}

In Fig.\,\ref{FRRcamp} we compare amplitude variability of all four
{\it Kepler} RRc stars. Only 150\,d of data is displayed. For the
consecutive objects (top to bottom) we adopted $\Delta t=\!10$\,d,
5\,d, 5\,d and 3\,d, respectively. Strong amplitude variations of
secondary periodicities are found in all the stars. These variations
are always irregular. As such, they are {\it not compatible with
beating} of two or more stable modes. The timescale of amplitude
changes is not the same in every RRc variable. It ranges from
$\sim\! 200$\,d (KIC\,5520878) to $\sim$10\,d (KIC\,9453114)
and becomes progressively shorter as we go from shorter to longer
period pulsators. This tendency explains why Fourier peaks in
Fig.\,\ref{FRRcprew} become broader with the increasing pulsation
period of the star.

As in the case of KIC\,5520878, secondary periodicities in all RRc
stars display significant phase variations (not shown). They occur
on the same timescale as the variations of the respective
amplitudes. We recall here that a phase change of a mode is
equivalent to a change of its frequency. Indeed, the difference
between the instantaneous and the mean frequency is

\begin{equation}
\Delta\omega = 2\pi\Delta f = \frac{d\phi}{dt}.
\label{EPervar}
\end{equation}

\noindent Thus, the frequencies of the secondary mode and its
subharmonics are not constant in {\it Kepler} RRc stars, but
fluctuate on a timescale of $10-200$\,d. This is the reason why
the values of $f_3/f_2$ and $f_{3'}/f_2$ (see Table\,\ref{TRRcfreq})
deviate from the exact half-integer ratios. These deviations are
larger in stars in which phase variations are faster.

\subsubsection{Frequency domain}\label{Sf2freq}

\begin{figure}
\vskip 1.9truecm
\centering
\resizebox{\hsize}{!}{\includegraphics{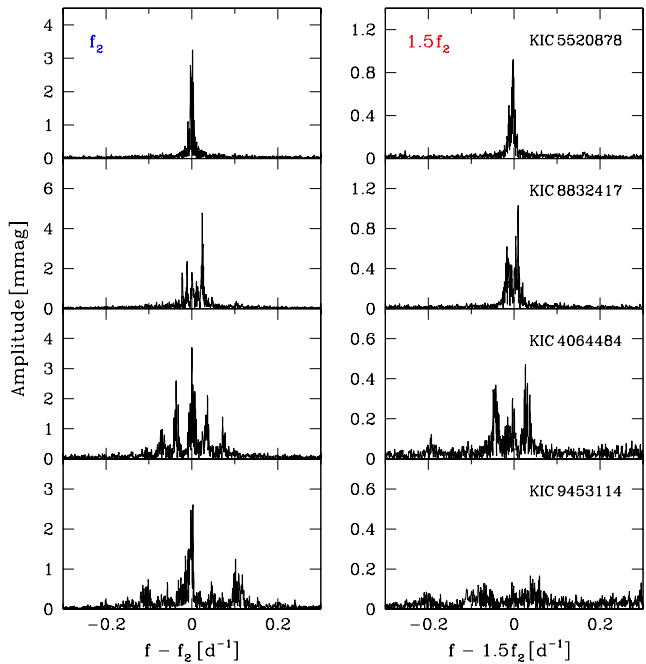}}
\caption{Prewhitened frequency spectra of {\it Kepler} RRc stars,
         computed for the entire light curves (Q0\,--\,Q10). Left
         column: FT of the secondary mode, $f_2$. Right column: FT
         of its subharmonic at $3/2f_2$.}
\label{Ff2f3}
\end{figure}

For each studied RRc star we computed the Fourier transform of the
entire light curve (Q0\,--\,Q10), after prewhitening it with the
dominant (radial) frequency and its harmonics. The resulting
frequency patterns in the vicinity of the secondary mode, $f_2$, and
its subharmonic, $3/2f_2$, are displayed in Fig.\,\ref{Ff2f3}.

Because the secondary mode in RRc stars has variable amplitude and
phase, it cannot be represented in the FT by a single sharp peak.
Fig.\,\ref{Ff2f3} confirms this. In the case of KIC\,5520878, $f_2$
is visibly broadened, but still does not split into resolved
components. For the other three stars, the mode splits into a
quintuplet of well-separated, equally-spaced peaks (see also
Fig.\,\ref{Ff2f1}). Such a pattern suggests that $f_2$ might
correspond to a multiplet of a $\ell=2$ nonradial mode. However,
this is not the only possible interpretation.

We notice that each component of the quintuplet is broadened and in
fact forms a band of power. None of them can be attributed to a
stable mode with a well-defined amplitude and frequency. We can
interpret this in two different ways. The $f_2$ frequency pattern
might be explained as a rotationally split multiplet of nonradial
modes, of the same $\ell$ and different $m$, all of which are
nonstationary. Alternatively, the observed pattern might result from
a quasi-periodic modulation of a single mode. In the latter picture,
the existence of a quasi-period is responsible for splitting $f_2$
into well-separated components (Benk\H{o}, Szab\'o \& Papar\'o 2011)
and the irregularity of the modulation causes these components to be
broadened. In Section\,\ref{Sdisvar} we will present arguments in
favour of this latter interpretation.

\begin{figure*}
\vskip 0.45truecm
\centering
\resizebox{\hsize}{!}{\includegraphics{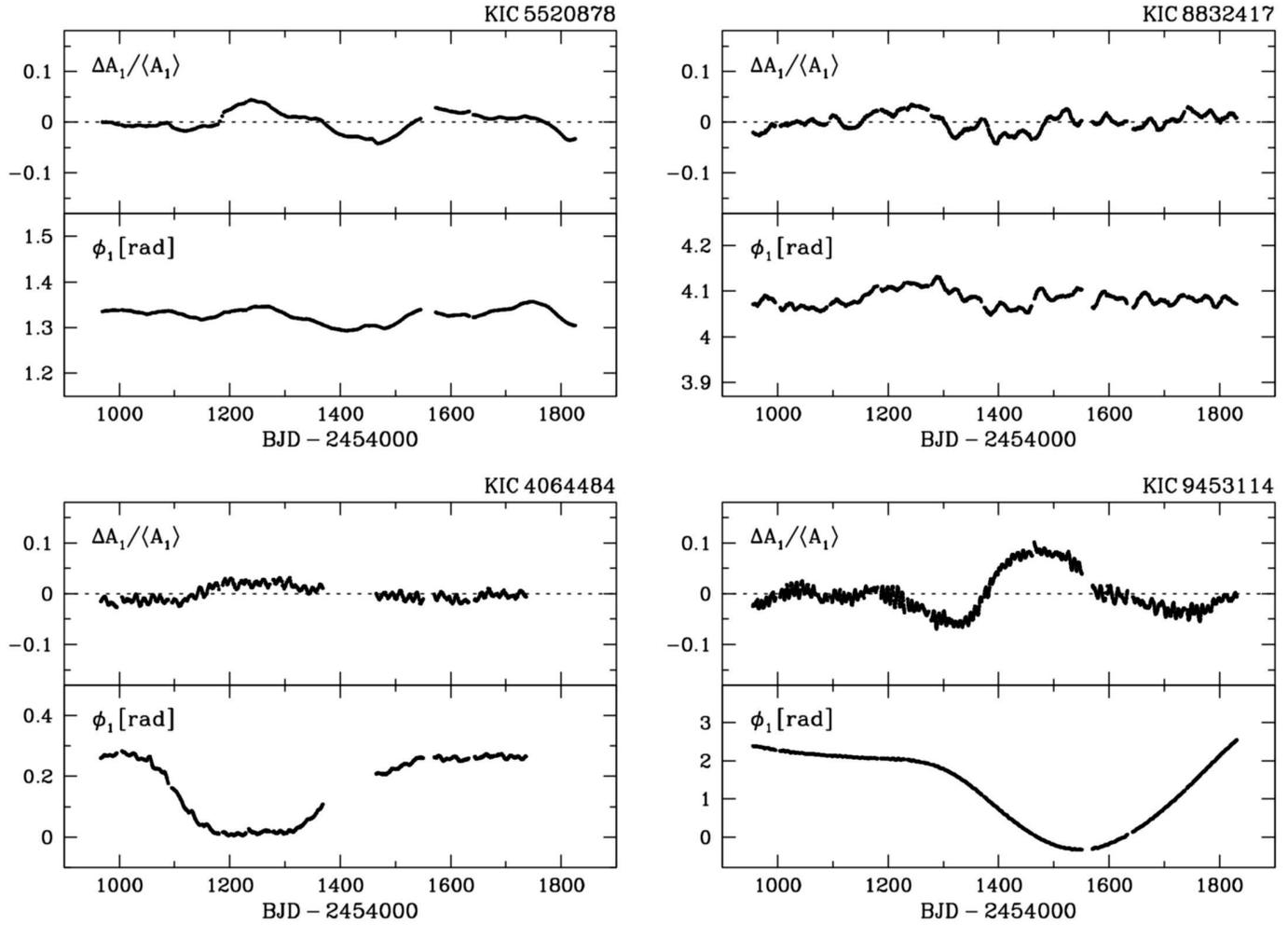}}
\caption{Amplitude and phase variations of the dominant radial mode
         in {\it Kepler} RRc stars. {\it This is low resolution
         version of the figure.}}
\label{FRRcA1}
\end{figure*}

The separation between the components of the $f_2$ multiplet can be
used to estimate the timescale of variability of this mode. For
KIC\,8832417 and KIC\,4064484 we find $\sim$\,75\,d and
$\sim$\,29\,d, respectively. In KIC\,9453114 the separation implies
timescale of $\sim$19\,d. However, in this object every second
multiplet component is very weak. Therefore, variations of $f_2$
will be dominated by a timescale half as long, i.e. $\sim$\,9.5\,d.
The derived numbers confirm our earlier assessment that the
variability of the secondary mode is faster in stars of longer
periods.

In the right column of Fig.\,\ref{Ff2f3} we display the Fourier
transform of the subharmonic, $3/2f_2$. This signal is visible only
in the FT of the first three stars. In KIC\,5520878 the peak is
broadened, but not yet resolved into separate components. This is
the same behaviour as displayed by its parent mode ($f_2$). In
KIC\,8832417 the subharmonic is already split, but because
components are much broader than in the case of  $f_2$, the
splitting appears incomplete. Finally, in KIC\,4064484 the
subharmonic is resolved into three separate bands of power. In both
KIC\,8832417 and KIC\,4064484 the splitting pattern of the
subharmonic looks different than for its parent mode. Nevertheless,
the {\it frequency separations} of multiplet components for $3/2f_2$
and for $f_2$ are roughly the same. This shows that both signals
vary on approximately same timescale.

\subsection{The radial mode}\label{}

\subsubsection{Time domain}\label{}

In Fig.\,\ref{FRRcA1} we show the temporal behaviour of the dominant
radial mode in {\it Kepler} RRc stars. In every object we find
changes of both the amplitude ($A_1$) and the phase ($\phi_1$). This
variability has two components: a long-term drift with a timescale
of many months, and a much faster quasi-periodic modulation which is
superimposed on this drift.

With the exception of KIC\,9453114, the long-term amplitude changes
are rather small, less than $\pm$\,4~per~cent. We are not sure if
these changes are real. They might be of instrumental origin,
resulting, e.g., from a slow image motion, coupled with a
contamination by a nearby faint star. Amplitude fluctuations of this
size are found in many stars observed with the {\it Kepler}
telescope. On the other hand, the amplitude variations found in
KIC\,9453114 are much larger ($\pm$\,8~per~cent) and they are almost
certainly intrinsic to the star. We note that these variations do
not resemble the Blazhko modulation as we currently know it
(Benk\H{o} et al. 2010, 2014).

In two of the studied stars the radial mode displays a large,
long-term drift of the pulsation phase, amounting to almost 0.3\,rad
in KIC\,4064484 and 2.9\,rad in KIC\,9453114. These variations are
orders of magnitude too fast to be explained by stellar evolution.
We note that nonevolutionary phase changes are observed in many RRc
and RRab stars (e.g. Jurcsik et al. 2001, 2012, Le\,Borgne et al.
2007), as well as in many overtone Cepheids (e.g. Berdnikov et al.
1997; Moskalik \& Ko{\l}aczkowski 2009). They are also occasionally
detected in $\delta$~Sct stars (e.g. Bowman \& Kurtz 2014).
Currently, their nature remains unexplained (see however Derekas et
al. 2004).

\begin{figure}
\vskip 0.25truecm
\centering
\resizebox{\hsize}{!}{\includegraphics{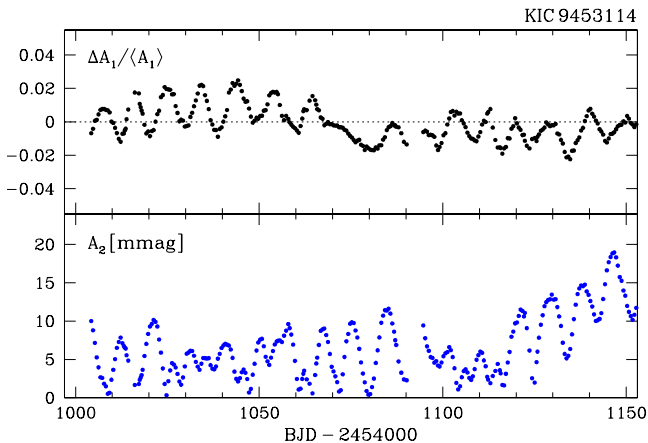}}
\caption{Amplitude variations in KIC\,9453114. Upper panel: dominant
         radial mode, $f_1$. Bottom panel: secondary mode, $f_2$.}
\label{FKIC94amp}
\end{figure}

We now turn our attention to the short timescale variations. They
are clearly visible in three of the {\it Kepler} RRc stars:
KIC\,8832417, KIC\,4064484 and KIC\,9453114. These variations are
very small: changes of $A_1$ are of the order of $\pm 1.5$~per~cent, or
equivalently of $\pm\, 2.5$\,mmag, and changes of $\phi_1$ of the
order of $\pm\, 0.01$\,rad. Such a small effect could be detected
only with ultraprecise photometry, such as delivered by {\it
Kepler}. The rapid variability of the radial mode is faster in stars
of longer periods. We recall that the same tendency was found
previously for the secondary mode, $f_2$. In Fig.\,\ref{FKIC94amp}
we compare temporal behaviour of the two modes in KIC\,9453114. It
is immediately obvious that the amplitude of the radial mode, $A_1$,
and the amplitude of the secondary mode, $A_2$, vary with the same
timescale. We also note that the two amplitudes seem to be
anticorrelated and a maximum (minimum) of one of them in most cases
coincides with a minimum (maximum) of the other. This behaviour is
common to all {\it Kepler} RRc stars.

\begin{figure}
\vskip 0.25truecm \centering
\resizebox{\hsize}{!}{\includegraphics{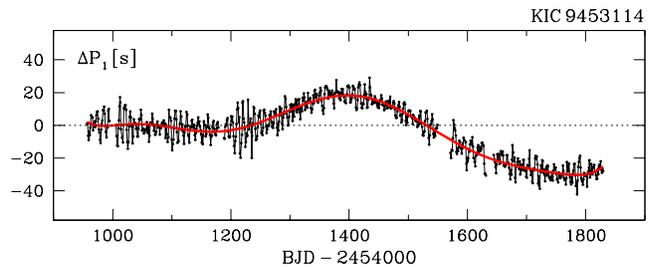}}
\caption{Period variations of the dominant radial mode in
         KIC\,9453114. Fifth order polynomial fit (red thick line)
         represents the slow period drift.}
\label{Pdot}
\end{figure}

\paragraph{Period variations}\label{}

Phase variations of the dominant radial mode (Fig.\,\ref{FRRcA1})
are equivalent to variations of its period, $P_1$. The latter can be
computed with Eq.\,\ref{EPervar}. Slow long-term phase drifts
observed in KIC\,4064484 and KIC\,9453114 correspond to slow
long-term changes of $P_1$, with a total range of $\sim$\,7\,s and
$\sim$\,49\,s, respectively. The short timescale phase variations
result in additional quasi-periodic modulation of $P_1$, imposed on
the long-term trends. The maximum range of these rapid variations is
$\pm$\,3.2\,s in KIC\,8832417, $\pm$\,5.0\,s in KIC\,4064484 and
$\pm$\,16.5\,s in KIC\,9453114. Changes of the radial mode period
in the last object are displayed in Fig.\,\ref{Pdot}.

\subsubsection{Frequency domain}\label{Sf1freq}

\begin{figure*}
\vskip 2.95truecm \centering
\resizebox{\hsize}{!}{\includegraphics{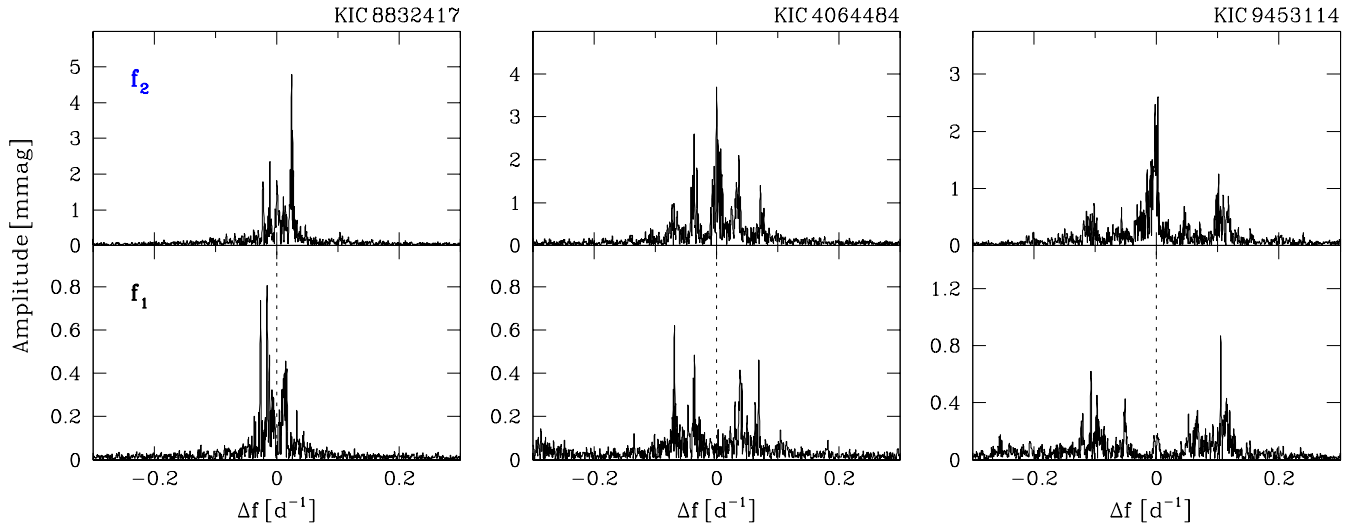}}
\caption{Frequency splitting for primary and secondary modes of {\it
         Kepler} RRc stars (for entire Q0\,--\,Q10 light curves).
         Upper panels: FT of the secondary mode, $f_2$. Bottom
         panels: FT of the dominant radial mode, $f_1$. The central
         component of the $f_1$ multiplet (indicated by the dashed
         line) has been subtracted.}
\label{Ff2f1}
\end{figure*}

Because of quasi-periodic modulation, we expect the dominant radial
mode to appear in the FT as a multiplet structured around the
frequency $f_1$. Since the variations are very small ($\sim$1.5~per~cent in
amplitude), the modulation sidepeaks should be {\it much lower} than
the central peak of the multiplet. Therefore, in order to extract
them from the FT, we first need to remove the central peak.

This is not a straightforward task. The amplitude $A_1$ and phase
$\phi_1$ of the radial mode undergo slow long-term changes
(Fig.\,\ref{FRRcA1}). As a result, the central component of the
multiplet is not coherent. The standard prewhitening method will
fail in such a case, leaving unremoved residual power. To remedy
this situation, we applied the {\it time-dependent prewhitening}, a
new technique described in Appendix\,\ref{Sappendix1}. In this
procedure, we subtract from the light curve a sine wave with varying
amplitude and phase. The functional form of $A_1(t)$ and $\phi_1(t)$
is determined by the time-dependent Fourier analysis of the data.
With the proper choice of the length of the light curve segment,
$\Delta t$, we can remove the central peak of the multiplet, leaving
the sidepeaks unaffected. The appropriate value of $\Delta t$ has to
be longer than the quasi-period of the modulation, but short enough
to capture the long-term drift of $A_1$ and $\phi_1$.

We applied this technique to all {\it Kepler} RRc stars except
KIC\,5520878, for which the radial mode does not show any short
timescale variations. In Fig.\,\ref{Ff2f1} (bottom panels) we
display the prewhitened FT of the radial mode in KIC\,8832417,
KIC\,4064484 and KIC\,9453114. For removing the central peak of the
multiplet we adopted $\Delta t = 100$\,d, 30\,d and 20\,d,
respectively. In the upper panels of the Figure we plot for
comparison the frequency pattern of the secondary mode, $f_2$. In
each of the three stars, the dominant radial mode, $f_1$, splits
into a quintuplet of equally spaced peaks. This is particularly well
visible for KIC\,4064484, where the structure is nicely resolved and
very clean. In KIC\,9453114 the quintuplet pattern is also clear,
but additional peaks appearing in the vicinity of $\Delta f =
-\,0.10$\,d$^{-1}$ and $\Delta f = +\,0.07$\,d$^{-1}$ slightly
confuse the picture. Compared to $f_2$, the modulation side peaks of
the radial mode are significantly narrower (i.e. more coherent).
They are also much lower, never exceeding 0.9\,mmag. This latter
property bears witness to the extremely low amplitude of the radial
mode modulation. Apart from these differences, the quintuplet
splitting patterns of $f_1$ and $f_2$ are very similar. In
particular, while separation between the quintuplet components
differs from star to star, it is always {\it the same for both
modes}. This means that the secondary mode, $f_2$, and the dominant
radial mode, $f_1$, are both modulated with a common timescale.

\section{Additional frequencies}\label{Sadditional}

\begin{figure*}
\vskip 4.4truecm
\centering
\resizebox{\hsize}{!}{\includegraphics{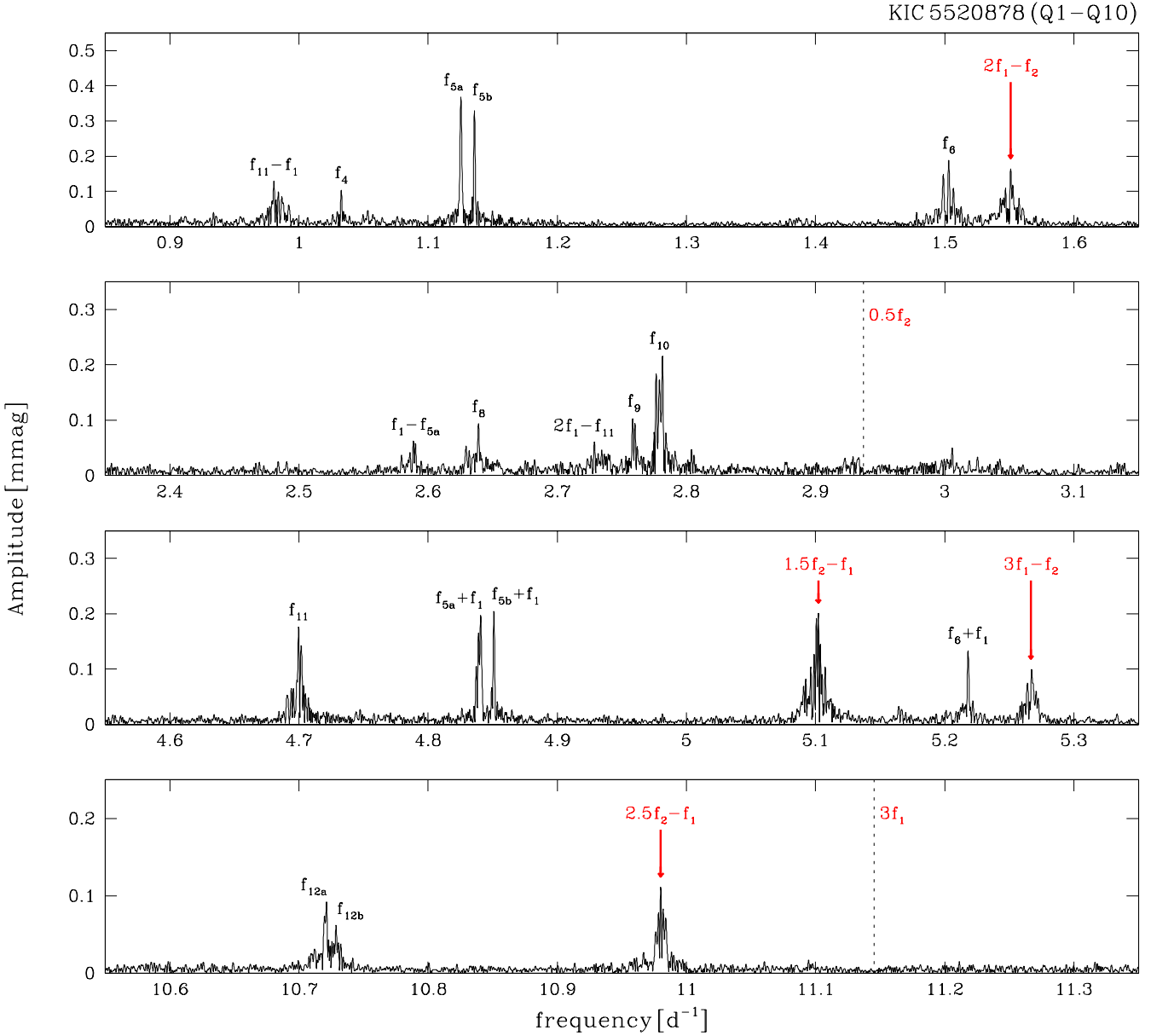}}
\caption{Additional low amplitude frequencies in KIC\,5520878.
         Frequency spectrum was computed for the entire light curve
         (Q1\,--\,Q10), prewhitend of the dominant radial mode, its
         harmonics and all peaks higher than 0.4\,mmag. Dashed lines
         indicate subtracted frequencies $1/2f_2$ and $3f_1$.}
\label{FKIC55add}
\end{figure*}

A Fourier transform of the entire light curve (Q0\,--\,Q10) offers not
only better frequency resolution, but also a much lower noise
level than the FT computed from a single quarter  of data
(Figs.\,\ref{FKIC55prew}, \ref{FRRcprew}). This advantage can be used
to search for additional, low amplitude periodicities hidden in the
light curve.

\subsection{KIC\,5520878}\label{}

\begin{table}
\vskip -0.1cm
\caption{Low-amplitude periodicities in KIC\,5520878}
\label{TKIC55add}
\centering
\begin{tabular}{lcc}
\hline
frequency          & $f$\,[d$^{-1}$]  & $A$\,[mmag] \\
\hline
$f_4$              & $\,$~1.03272     & 0.11        \\
\noalign{\smallskip}
$f_{5a}$           & $\,$~1.12535     & 0.35        \\
$f_{5b}$           & $\,$~1.13577     & 0.33        \\
$f_{5a}-f_1$       & $\,$~2.59012     & 0.07        \\
$f_{5b}\;\!-f_1$   &      2.5793      & 0.04        \\
$f_{5a}-2f_1$      &      6.3049      & 0.04        \\
$f_{5b}\;\!-2f_1$  &      6.2941      & 0.03        \\
$f_{5a}-3f_1$      &     10.0201~$\,$ & 0.04        \\
$f_{5b}\;\!-3f_1$  &     10.0094~$\,$ & 0.03        \\
$f_{5a}-4f_1$      &     13.7353~$\,$ & 0.03        \\
$f_{5b}\;\!-4f_1$  &     13.7245~$\,$ & 0.03        \\
$f_{5a}+f_1$       & $\,$~4.84070     & 0.20        \\
$f_{5b}\;\!+f_1$   & $\,$~4.85087     & 0.21        \\
$f_{5a}+2f_1$      &      8.5558      & 0.04        \\
$f_{5b}\;\!+2f_1$  & $\,$~8.56611     & 0.06        \\
$f_{5a}+3f_1$      &     12.2709~$\,$ & 0.03        \\
$f_{5b}\;\!+3f_1$  &     12.2813~$\,$ & 0.03        \\
$f_{5a}+4f_1$      &     15.9859~$\,$ & 0.03        \\
\noalign{\smallskip}
$f_{6a}$           & $\,$~1.49877     & 0.14        \\
$f_{6b}$           & $\,$~1.50317     & 0.19        \\
$f_{6c}$           & $\,$~1.50686     & 0.11        \\
$f_{6b}\;\!+f_1$   & $\,$~5.21814     & 0.13        \\
$f_{6b}\;\!+2f_1$  &      8.9332      & 0.05        \\
$f_{6b}\;\!+3f_1$  &     12.6483~$\,$ & 0.03        \\
\noalign{\smallskip}
$f_7$              & $\,$~1.80060     & 0.06        \\
\noalign{\smallskip}
$f_8$              & $\,$~2.63891     & 0.08        \\
\noalign{\smallskip}
$f_9$              & $\,$~2.75926     & 0.10        \\ 
$f_9+f_1$          &      6.4744      & 0.04        \\
\noalign{\smallskip}
$f_{10a}$          & $\,$~2.77651     & 0.19        \\
$f_{10b}$          & $\,$~2.77900     & 0.19        \\ 
$f_{10c}$          & $\,$~2.78141     & 0.22        \\
$f_{10a}+f_1$      & $\,$~6.49165     & 0.08        \\
$f_{10b}\;\!+f_1$  & $\,$~6.49419     & 0.08        \\
$f_{10c}\;\!+f_1$  & $\,$~6.49645     & 0.08        \\
$f_{10a}+2f_1$     &     10.2067~$\,$ & 0.03        \\
\noalign{\smallskip}
$f_{11}$           & $\,$~4.69968     & 0.18        \\ 
$f_{11}-f_1$       & $\,$~0.98052     & 0.14        \\
$f_{11}-2f_1$      & $\,$~2.72874     & 0.06        \\
$f_{11}+f_1$       & $\,$~8.41479     & 0.05        \\
$f_{11}+2f_1$      &     12.1298~$\,$ & 0.04        \\
\noalign{\smallskip}
$f_{12a}$          &     10.72126     & 0.09        \\
$f_{12b}$          &     10.72867     & 0.07        \\
$f_{12a}-f_1$      &      7.0061      & 0.05        \\
$f_{12a}+f_1$      &     14.4363~$\,$ & 0.04        \\
$f_{12a}+2f_1$     &     18.1516~$\,$ & 0.03        \\
$f_{12b}\;\!+2f_1$ &     18.1586~$\,$ & 0.02        \\
\hline
\end{tabular}
\end{table}


\begin{table}
\vskip -0.1cm
\caption{Low-amplitude periodicities in KIC\,8832417}
\label{TKIC88add}
\centering
\begin{tabular}{lcc}
\hline
frequency & $f$\,[d$^{-1}$] & $A$\,[mmag] \\
\hline
$f_4$     & $\,$~1.18653    & 0.25        \\
$f_4+f_1$ & $\,$~5.20990    & 0.12        \\
\noalign{\smallskip}
$f_{5a}$  & $\,$~3.08589    & 0.07        \\
$f_{5b}$  & $\,$~3.09347    & 0.08        \\
\noalign{\smallskip}
$f_6$     & $\,$~6.65658    & 0.32        \\ 
$f_6-f_1$ & $\,$~2.63315    & 0.10        \\
$f_6+f_1$ &     10.67988    & 0.11        \\
\noalign{\smallskip}
$f_7$     & $\,$~9.51594    & 0.08        \\
\hline
\end{tabular}
\end{table}


\begin{table}
\vskip -0.1cm
\caption{Low-amplitude periodicities in KIC\,9453114}
\label{TKIC94add}
\centering
\begin{tabular}{lcc}
\hline
frequency  & $f$\,[d$^{-1}$] & $A$\,[mmag] \\
\hline
$f_4$      & 0.91199         & 0.23        \\
$f_4+f_1$  & 3.64315         & 0.10        \\
\noalign{\smallskip}
$f_5$      & 1.87588         & 0.92        \\
$f_5-f_1$  & 0.85617         & 0.10        \\
$f_5-2f_1$ & 3.58750         & 0.16        \\
\hline
\end{tabular}
\end{table}



In order to lower the noise in the FT as much as possible, we first
subtracted from the light curve the dominant frequency $f_1$ and its
harmonics, as well as eight other frequency components higher than
0.4\,mmag ($f_2$, $f_2$\,--\,$f_1$, $f_1$\,+\,$f_2$,
$2f_1$\,+\,$f_2$, $1/2f_2$, $3/2f_2$, $f_1-1/2f_2$, $f_1+3/2f_2$).
This step was performed with the time-dependent prewhitening method,
adopting $\Delta t = 20$\,d. The prewhitened FT yields a rich
harvest of additional low-amplitude periodicities. We list them in
Table\,\ref{TKIC55add}. Four segments of the prewhitened FT are
displayed in Fig.\,\ref{FKIC55add}.

In total, we detected 46 new, significant frequencies. All have
amplitudes below 0.4\,mmag. We checked that none of them can be
explained as a linear combination of $f_1$, $f_2$ or $1/2f_2$. Only
15 of the new frequencies are independent, i.e., correspond to
pulsation modes. The remaining frequencies are linear combinations
of the independent ones and $f_1$. In fact, except of $f_4$, $f_7$
and $f_8$, all other independent frequencies generate at least one
combination peak. This is a very important observation. It proves
that these periodicities originate in KIC\,5520878 itself, and are
not introduced by blending with another star.

The strongest of the low-amplitude modes appear at $f_{5a} =
1.12535$\,d$^{-1}$ and $f_{5b} = 1.13577$\,d$^{-1}$. These two modes
form a well-resolved doublet. The same doublet structure is apparent
in their linear combinations with $f_1$. Another well-resolved
doublet with slightly smaller separation is formed by $f_{12a} =
10.72126$\,d$^{-1}$ and $f_{12b} = 10.72867$\,d$^{-1}$. Two
equidistant triplets are also found: $f_6$ with separation of
$\delta f_6 = 0.00405$\,d$^{-1}$ and $f_{10}$ with separation of
$\delta f_{10} = 0.00245$\,d$^{-1}$. With the timebase of 869\,d, we
consider the $f_{10}$ triplet to be only marginally resolved. We
note that $\delta f_{10}$ is not very different from $1/372.5\,{\rm
d} = 0.00268$\,d$^{-1}$. Therefore, splitting of $f_{10}$ might be
an artefact, caused by the instrumental amplitude variation with the
orbital period of the telescope. We think this is unlikely. If such
a modulation is present in the data, it should affect {\it all} the
frequencies. Clearly, this is not the case. Judging from behaviour
of the dominant radial mode (Fig.\,\ref{FRRcA1}), any instrumental
amplitude variation in the KIC\,5520878 dataset cannot be larger
than $\pm 4$ per cent. Such a small modulation cannot explain
relatively large sidepeaks of the $f_{10}$ triplet.

Two modes attract special attention. For $f_9$ and $f_{10}$ we find
period ratios of $P_1/P_9 = 0.7427$ and $P_1/P_{10} = 0.7480$,
respectively. These values are close to canonical period ratio of
the radial first overtone and radial fundamental mode in the
RR~Lyrae stars (Kov\'acs 2001; Moskalik 2014). This suggests that
either $f_9$ or $f_{10}$ might correspond to the radial fundamental
mode. However, such a conclusion would be too hasty. In the case of
the RR~Lyrae stars, there is an empirical relation between the
period ratio of the two lowest radial modes, $P_{\rm 1O}/P_{\rm F}$
and the pulsation period itself. $P_{\rm 1O}/P_{\rm F}$ is a steep
function of $P_{\rm F}$ and at short periods it becomes considerably
lower than 0.74. This is best demonstrated by the double-mode
RR~Lyrae pulsators (RRd stars) of the Galactic Bulge (Soszy\'nski et
al. 2011a). When plotted on the Petersen diagram, i.e. period ratio
vs. period diagram (their Fig.\,4), the Bulge RRd stars form a
narrow, well defined sequence, which extends down to $P_{\rm
1O}/P_{\rm F} \!\sim\! 0.726$ at a fundamental-mode period of
$P_{\rm F} \!\sim\! 0.347$\,d. It is easy to check that neither
$P_1/P_9$ nor $P_1/P_{10}$ conforms to this empirical progression.
Using Fig.\,4 of Soszy\'nski et al. (2011a) we can estimate that the
radial fundamental mode in KIC\,5520878 should have a period of
$P_{\rm F} = 0.3677-0.3692$\,d ($f_{\rm F} =
2.7084-2.7194$\,d$^{-1}$), corresponding to a period ratio of
$P_{\rm 1O}/P_{\rm F} \equiv P_1/P_{\rm F} = 0.729-0.732$. We
conclude, therefore,  that none of the frequencies detected in
KIC\,5520878 can be identified with the radial fundamental mode.

Another frequency of special interest is $f_{11}$. In the case of
this mode we find a period ratio of $P_{11}/P_1 \equiv P_{11}/P_{\rm
1O} = 0.7905$. This value is close to period ratios measured in
double overtone Cepheids, particularly in those of the Galactic
Bulge and the Galactic field (Soszy\'nski et al. 2008, 2010, 2011b;
Smolec \& Moskalik 2010 and references therein). We do not know
1O+2O pulsators among the RR~Lyrae stars, so we cannot compare their
period ratios with $P_{11}/P_1$. But we can proceed differently --
using the estimated period of the radial fundamental mode in
KIC\,5520878, we can estimate $P_{11}/P_{\rm F} = 0.576-0.579$. This
value is very close to the empirical period ratio of the F+2O
double-mode RR~Lyrae pulsators (Moskalik 2013). On this basis, we
identify $f_{11}$ with the radial second overtone.

\begin{table*}
\vskip -0.15cm
\caption{Known RR Lyrae variables with $P_2/P_1 \sim 0.61$}
\label{TRRcother}
\centering
\begin{tabular}{llccccc}
\hline
Star                   & Type   & $P_1$  & $P_2/P_1$ & $A_2$   & subharmonics & ref. \\
                       &        & [d]    &           & [mmag]  & of $f_2$     &      \\
\hline
$\omega$~Cen V10       & \,RRc  & 0.3750 & 0.6137    & \,~6.4  &              & 2    \\
$\omega$~Cen V19       & \,RRc  & 0.3000 & 0.6119    & \,~7.1  &              & 2    \\
$\omega$~Cen V81       & \,RRc  & 0.3894 & 0.6138    & \,~7.2  &              & 2    \\
$\omega$~Cen V87       & \,RRc  & 0.3965 & 0.6219    & \,~6.3  &              & 2    \\
$\omega$~Cen V105      & \,RRc  & 0.3353 & 0.6138    &   12.5  &              & 2    \\
$\omega$~Cen V350      & \,RRc  & 0.3791 & 0.6084    & \,~6.2  &              & 2    \\
\noalign{\smallskip}
OGLE-LMC-RRLYR-11983   & \,RRc  & 0.3402 & 0.6026    &  ~51.5: &              & 3    \\
OGLE-LMC-RRLYR-14178   & \,RRc  & 0.3634 & 0.6103    &  ~27.8: &              & 3    \\
\noalign{\smallskip}
SDSS Stripe 82-1528004 & \,RRc: & 0.3276 & 0.6068    &    ?    &              & 4    \\
SDSS Stripe 82-3252839 & \,RRc  & 0.3112 & 0.6238    &    ?    &              & 4    \\
\noalign{\smallskip}
CoRoT 0105036241       & \,RRc  & 0.3729 & 0.6125    & \,~3.3  &              & 6    \\
CoRoT 0105735652       & \,RRc  & 0.2792 & 0.6150    & \,~2.2  &              & 6    \\
\noalign{\smallskip}
KIC 4064484            & \,RRc  & 0.3370 & 0.6156    & \,~4.6  &  yes         & 7    \\
KIC 5520878            & \,RRc  & 0.2692 & 0.6320    & \,~7.0  &  yes         & 7    \\
KIC 8832417            & \,RRc  & 0.2485 & 0.6122    & \,~7.1  &  yes         & 7    \\
KIC 9453114            & \,RRc  & 0.3661 & 0.6144    & \,~4.6  &  yes         & 7    \\
\noalign{\smallskip}
EPIC 60018224          & \,RRc  & 0.3063 & 0.6145    &   11.4  &  yes         & 8    \\
EPIC 60018238          & \,RRc  & 0.2748 & 0.6030    & \,~2.0  &              & 8    \\
EPIC 60018678          & \,RRc  & 0.4325 & 0.6200    & \,~3.8  & (yes)        & 8    \\
\noalign{\medskip}
AQ Leo                 & \,RRd  & 0.4101 & 0.6211    & \,~2.5  &  yes         & 1    \\
CoRoT 0101368812       & \,RRd  & 0.3636 & 0.6141    & \,~5.5  &  yes         & 5    \\
EPIC 60018653          & \,RRd  & 0.4023 & 0.6163    & \,~8.5  &  yes         & 8    \\
EPIC 60018662          & \,RRd  & 0.4175 & 0.6170    & \,~6.9  &  yes         & 8    \\
\hline
\multicolumn{7}{p{12cm}}{\small REFERENCES: 1 - Gruberbauer et al. (2007);
2 - Olech \& Moskalik (2009); 3 - Soszy\'nski et al. (2009);
4 - S\"uveges et al. (2012); 5 - Chadid (2012); 6 - Szab\'o et al. (2014);
7 - this paper; 8 - Moln\'ar et al. in preparation.} \\
\end{tabular}
\end{table*}


Except for $f_{11}$, all the remaining low-amplitude modes detected
in KIC\,5520878 must be nonradial. We note that most of them ($f_4$,
$f_5$, $f_6$, $f_7$, $f_8$) have frequencies lower than the
(unobserved) radial fundamental mode. This implies that these are
not $p$-mode oscillations, they must be classified as gravity modes
($g$-modes). This is the first detection of such modes in RR~Lyrae
stars.

\subsection{KIC\,4064484, KIC\,8832417 and KIC\,9453114}\label{}

In the other three {\it Kepler} RRc stars only very few additional
low-amplitude modes can be found. In KIC\,8832417 and KIC\,9453114
we detected 5 and 2 such modes, respectively. Their frequencies are
listed in Tables\,\ref{TKIC88add} and \ref{TKIC94add}. About half of
them generate linear combinations with the dominant frequency,
$f_1$. None of the low-amplitude modes can be identified with the
radial fundamental or with the second radial overtone. As in
KIC\,5520878, some of the modes have frequencies lower than the
fundamental radial mode. Therefore, they must be $g$-mode type. This
is the case for $f_4$ in KIC\,8832417 and for both $f_4$ and $f_5$
in KIC\,9453114.

The strongest of the low-amplitude modes detected in KIC\,8832417 is
of particular interest. Its frequency $f_6 = 6.65658$\,d$^{-1}$
yields a period ratio of $P_6/P_1 = 0.6044$, which is very close to
$P_2/P_1 = 0.6122$. Apparently, in this star there are two modes
with period ratios in the range of $0.60-0.63$.

No frequencies beyond $f_1$, $f_2$, $3/2f_2$ and their combinations
were found in KIC\,4064484.

\section{Discussion}\label{Sdiscussion}

\subsection{A new period ratio for RR Lyrae stars}\label{}

It is striking that all four RRc stars prominent in the {\it Kepler}
field that have been analyzed in this paper show an additional
frequency with a period ratio of $\sim\!0.61$ to the main radial
mode. This is not the first time such a period ratio has been seen
in RR~Lyrae stars. It was first found in a double-mode variable
AQ~Leo (Gruberbauer et al. 2007). Since then, low amplitude modes
yielding similar period ratios have been detected in 18 other
RR~Lyrae variables, observed both from the ground and from space. We
list these objects in Table\,\ref{TRRcother}, where we also include,
for completeness, the four {\it Kepler} RRc stars. For each variable
we provide its pulsation type, a period of the dominant radial mode,
$P_1$, a period ratio of the secondary mode and the dominant mode,
$P_2/P_1$, and an amplitude of the secondary mode, $A_2$. In the
case of double-mode variables in which two radial modes are present,
$P_1$ refers to the {\it first radial overtone}. The amplitudes
$A_2$ are given in different photometric systems by different
authors and, consequently, are not directly comparable to each
other. We quote them here only to provide a rough estimate of the
strength of the secondary mode.

\medskip

\noindent\underline{Comments on individual stars}

\medskip

\noindent {\bf $\omega$~Cen variables V10, V81, V87 and V350:}
Detection of the secondary mode in these stars is unambiguous, but
identification of its true period is hindered by daily aliases.
Depending on the choice of an alias we find period ratio of either
$P_2/P_1\!\sim\! 0.80$ or $\sim\! 0.61$. Only in the case of V10 is
the former alias slightly higher, which led Olech \& Moskalik (2009)
to designate this variable a candidate double overtone pulsator. In
V81, V87 and V350 the latter alias is higher. So far, no unambiguous
1O+2O double-mode RR~Lyrae stars have been found in any stellar
system (see, however, Pigulski 2014). On the other hand, the
existence of RR~Lyrae stars with $P_2/P_1\!\sim\! 0.61$ is well
established. On these grounds, we consider the period ratio of
$\sim\! 0.61$ to be more likely in these four variables.

\smallskip

\noindent {\bf $\omega$~Cen variables V19 and V105: } In these stars
the identification of the correct alias and consequently of the true
period of the secondary mode is unambiguous.

\smallskip

\begin{figure}
\vskip 3.9truecm
\centering
\resizebox{\hsize}{!}{\includegraphics{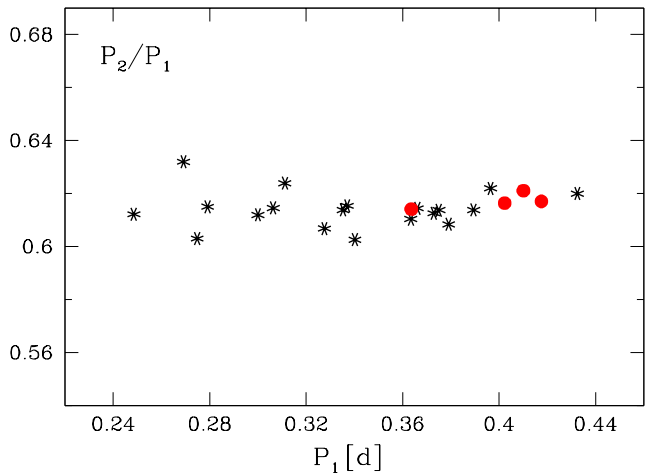}}
\caption{Petersen diagram for RR~Lyrae stars of
         Table\,\ref{TRRcother}. RRc stars are plotted with black
         asterisks and RRd stars with red filled circles.}
\label{fig13}
\end{figure}

\noindent {\bf OGLE-LMC-RRLYR-11983:} In the original paper of
Soszy\'nski et al. (2009) the star was classified as a fundamental
mode pulsator (RRab star). This was based on the measured value of
the Fourier phase $\phi_{21}$. However, the harmonic of the primary
mode in this variable is weak and its amplitude is determined with a
large error of almost 30~per~cent. Consequently, $\phi_{21}$ is not
measured accurately enough to distinguish between an RRc and an RRab
light curve. In fact, $\phi_{21} = 3.358$, derived by Soszy\'nski et
al. (2009), does not fit to either type. We have reclassified this
star using the empirical period luminosity relation for the
Wesenheit index, $W_I = I-1.55(V-I)$. With $W_I = 18.376$\,mag,
OGLE-LMC-RRLYR-11983 is placed firmly among the RRc stars of the
LMC.

\smallskip

\noindent {\bf OGLE-LMC-RRLYR-14178:} The star was classified by
Soszy\'nski et al. (2009) as an overtone pulsator (RRc star). The
value of the Wesenheit index $W_I = 17.795$\,mag supports this
classification.

\smallskip

\noindent {\bf SDSS Stripe 82 variables 1528004 and 3252839:} The
secondary frequency in these stars was detected with a principal
component analysis, applied to multiband photometric data (S\"uveges
et al. 2012). The authors did not identify the dominant radial mode.
On the basis of the empirical period-amplitude diagram (their
Fig.\,9) we classify variable 3252839 as an RRc star. In the case of
1528004, the period of the radial mode points towards overtone
pulsations, but its amplitude is too high for an RRc type. We
classify this variable as a probable RRc star. Amplitudes of the
secondary mode were not published.

\smallskip

\noindent {\bf CoRoT~0105036241 and CoRoT~0105735652:} The secondary
mode in these stars was detected with photometry collected by the
CoRoT space telescope (Szab\'o et al. 2014). No subharmonics were
found.

\smallskip

\noindent {\bf EPIC~60018224 $\equiv$ EV~Psc: } The secondary mode
was discovered during 9-d long engineering test run of {\it Kepler}
K2 mission (Moln\'ar et al. in preparation). Subharmonics of the
secondary mode at $\sim\! 1/2 f_2$ and $\sim\! 3/2 f_2$ are also
clearly visible.

\smallskip

\noindent {\bf EPIC~60018238:} No subharmonics of the secondary mode
were detected in this star.

\smallskip

\begin{figure}
\vskip 3.9truecm
\centering
\resizebox{\hsize}{!}{\includegraphics{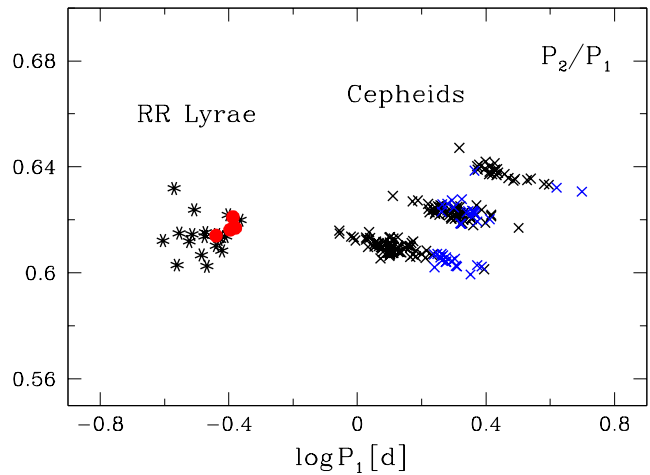}}
\caption{Petersen diagram for RR~Lyrae stars and Cepheids with
         period ratios of $0.60-0.65$. For the RR~Lyrae variables,
         the same symbols as in Fig.\,\ref{fig13} are used. Cepheids
         of SMC and LMC are plotted with black and with blue
         crosses, respectively.}
\label{fig14}
\end{figure}

\noindent {\bf EPIC~60018678:} In addition to the secondary mode,
marginally significant subharmonics at $\sim\! 1/2 f_2$ and $\sim\!
3/2 f_2$ are also present in this star (Moln\'ar et al. in
preparation).

\smallskip

\noindent {\bf AQ~Leo:} The secondary mode in this double-mode star
was discovered with photometry collected by the MOST space telescope
(Gruberbauer et al. 2007). A subharmonic of a secondary mode at
$\sim\! 1/2 f_2$ was also found, although it was not recognized as
such at the time.

\smallskip

\noindent {\bf CoRoT 0101368812: } The secondary mode with a period
ratio of $\sim\! 0.61$ to the first radial overtone was discovered
by Chadid (2012). In addition, its subharmonic at $\sim\! 3/2 f_2$
was also detected, but not recognized as such by the author. It is
denoted in the original paper as $f_4-f_1$.

\smallskip

\noindent {\bf EPIC 60018653:} In addition to the secondary mode,
subharmonics at $\sim\! 1/2 f_2$ and $\sim\! 3/2 f_2$ are also
clearly visible in this star (Moln\'ar et al. in preparation).

\smallskip

\noindent {\bf EPIC 60018662:} In addition to the secondary mode, a
weak subharmonic at $\sim\! 1/2 f_2$ was also detected in this star
(Moln\'ar et al. in preparation).

\medskip

Objects listed in Table\,\ref{TRRcother} form a sample of 23
RR~Lyrae variables in which a period ratio of $\sim\! 0.61$ has been
found. In most of these stars the main mode is the first radial
overtone (RRc type). Four variables belong to the group of
double-mode pulsators (RRd type), where two radial modes are
excited. Even in the latter case, the first radial overtone strongly
dominates, having an amplitude $1.8-3.0$ times higher than the
fundamental mode (Gruberbauer et al. 2007; Chadid 2012; Moln\'ar et
al. in preparation). In Fig.\,\ref{fig13} we plot all 23 stars on
the Petersen diagram. They form a tight, almost horizontal
progression, with values of $P_2/P_1$ restricted to a narrow range
of $0.602-0.632$. Both RRc and RRd variables follow the same trend.
Clearly, the RR~Lyrae star with $P_2/P_1\!\sim 0.61$ form a highly
homogenous group, constituting a {\it new, well defined class of
multimode pulsators}.

In all stars of this class the amplitude of the additional mode is
extremely low, in the mmag range. This amplitude is typically $20 -
60$ times lower than the amplitude of the dominant radial mode.
Detection of such a weak signal from the ground is difficult. The
situation is very different for stars observed from space. With the
sole exception of the Blazhko star CSS~J235742.1--015022 (Moln\'ar
et al. in preparation), the secondary mode yielding period ratio of
$P_2/P_1\!\sim 0.61$ has been detected in {\it every} RRc and RRd
pulsator for which high-precision space photometry is available.
This is 13 objects out of total 14 observed from space. Such
statistics strongly suggest that excitation of this additional mode
is not an exception. It must be a common property of RRc and RRd
variables. We expect that it should be found in many more stars,
provided that good enough data become available.

\subsubsection{Comparison with Cepheids}\label{}

Detection of low amplitude secondary modes with a period ratio of
$\sim\! 0.61$ to the main radial mode is not limited to the RR~Lyrae
stars. Similar modes are also found in Classical Cepheids of the
Magellanic Clouds. So far, 173 such variables have been identified
(Moskalik \& Ko{\l}aczkowski 2008, 2009; Soszy\'nski et al. 2008,
2010). In Fig.\,\ref{fig14} we plot them on the Petersen diagram,
together with their RR~Lyrae counterparts. Both groups have very
similar properties. In the case of Cepheids, just like in the
RR~Lyrae stars, the phenomenon occurs only in the first overtone
variables or in the double-mode variables pulsating simultaneously
in the fundamental mode and the first overtone (only one star).
Apparently, excitation of the first radial overtone is a necessary
condition. The amplitudes of the secondary modes are as low as in
the RR~Lyrae stars, with amplitude ratios of $A_2/A_1 < 0.055$
(Moskalik \& Ko{\l}aczkowski 2009). The measured period ratios are
also almost the same in both types of pulsators, although in
Cepheids the range of $P_2/P_1$ is somewhat broader ($0.599-0.647$).
The only difference between the two groups is that Cepheids split in
the Petersen diagram into three well-detached parallel sequences,
whereas no such structure is evident in the case of RR~Lyrae stars.
This difference needs to be explained by future theoretical work.

\subsubsection{Comparison with theoretical calculations}\label{}

In the case of Classical Cepheids it has been shown that the period
ratio of $\sim\! 0.61$ cannot be reproduced by two radial modes
(Dziembowski \& Smolec 2009; Dziembowski 2012). Since the main mode
is radial, this implies that the secondary frequency, $f_2$, must
correspond to a nonradial mode of oscillation. Theoretical analysis
indicates that an f-mode of high spherical degree ($\ell=42-50$) is
the most likely candidate (Dziembowski 2012).

\begin{figure}
\vskip 3.9truecm
\centering
\resizebox{\hsize}{!}{\includegraphics{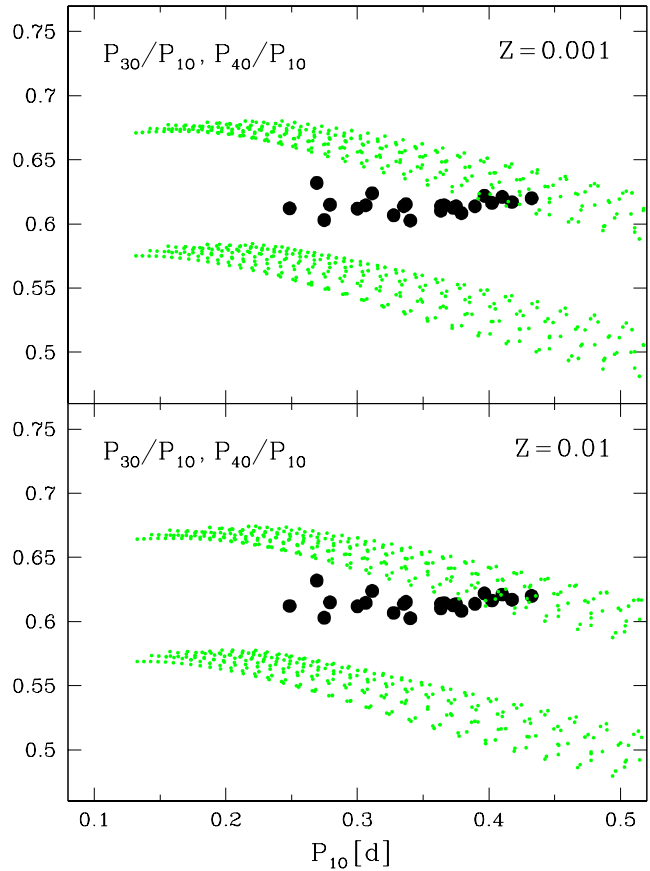}}
\caption{Linear period ratios $P_{\rm 3O}/P_{\rm 1O}$ and $P_{\rm
         4O}/P_{\rm 1O}$. Model masses and luminosities are in the
         range of $0.55-0.75$\,${\rm M}_\odot$ and $40-70$\,${\rm
         L}_\odot$, respectively. Upper panel: models for
         metallicity of $Z=0.001$. Bottom panel: models for
         metallicity of $Z=0.01$. The RR~Lyrae stars of
         Table\,\ref{TRRcother} are plotted with filled circles for
         comparison.}
\label{fig15}
\end{figure}

The same argument for nonradial nature of $f_2$ can also be made for
RR~Lyrae stars. In Fig.\,\ref{fig15} we present theoretical Petersen
diagrams for two different metallicities, $Z\!=\!0.001$ and
$Z=0.01$. Period ratios of the third-to-first ($P_{\rm 3O}/P_{\rm
1O}$) and fourth-to-first ($P_{\rm 4O}/P_{\rm 1O}$) radial overtones
have been computed with the Warsaw linear pulsation code (Smolec \&
Moskalik 2008a). Our calculations cover a range of masses,
luminosities and effective temperatures spanning the entire domain
of the RR~Lyrae stars. For metallicity $Z\! =\! 0.01$ the
theoretical period ratios are systematically lower than for $Z\! =\!
0.001$, but the difference is not large. Further increase of $Z$
does not lower the computed values any further. We note in passing,
that weak sensitivity to $Z$ is typical of period ratios of two
overtones. This is different from behaviour of $P_{\rm 1O}/P_{\rm
F}$, sensitivity of which to a metal abundance is much stronger (see
e.g. Popielski et al. 2000). The models are compared with the RRc
and RRd variables listed in Table\,\ref{TRRcother}. The observed
period ratios generally fall between theoretically predicted values
of $P_{\rm 3O}/P_{\rm 1O}$ and $P_{\rm 4O}/P_{\rm 1O}$. Only in a
handful of long period variables can the secondary period be matched
with the third overtone, but for vast majority of the sample it
cannot be matched with a radial mode. This result does not depend on
the choice of $Z$. Therefore, as in the case of Cepheids, we
conclude that also in the RR~Lyrae stars the secondary mode with a
period ratio of $\sim\! 0.61$ to the first radial overtone must be
nonradial.

Model calculations show that two different types of nonradial modes
are linearly unstable in the RR~Lyrae stars (Van Hoolst et al. 1998;
Dziembowski \& Cassisi 1999). Low-degree modes of $\ell=1\,-\,3$ are
excited rather weakly and preferentially in the vicinity of radial
modes. Their frequency spectrum is very dense, which makes it
difficult to destabilize only a single isolated mode. The other
group are strongly trapped unstable modes (STU modes) of high
spherical degree ($\ell \ge 6$). Oscillations of this type are
excited strongly, with growth rates as large as in the case of radial
modes. Most importantly, such oscillations can be excited at
frequencies far apart from those of the radial modes. The strong
trapping in the envelope, which makes their destabilization
possible, usually selects from the dense spectrum only a single
mode. We believe that excitation of STU modes is the most likely
explanation for the puzzling period ratio of $\sim~\!\! 0.61$ in the
RRc and RRd variables. At this point, this is only a working
hypothesis. Its verification requires detailed linear nonadiabatic
pulsation calculations, similar to those performed by Dziembowski
(2012) for the Classical Cepheids.

\subsection{Period doubling of the secondary mode}\label{}

In all four {\it Kepler} RRc stars not only do we see the additional
frequency with a period ratio of $\sim\! 0.61$ to the main mode, but
we also see its subharmonics at $\sim 1/2 f_2$ and $\sim 3/2 f_2$.
Subharmonic frequencies are also detected in several other variables
of Table\,\ref{TRRcother}. Noticeably, they are found only in those
stars for which high precision space photometry is available. This
is hardly surprising, considering that amplitudes of the
subharmonics are even lower than in the case of $f_2$, and almost never
exceed 3.0\,mmag. In total, the additional mode $f_2$ is accompanied
by subharmonics in 10 out of 13 RRc and RRd stars studied from
space. Clearly, this is a common property of these variables.

The presence of subharmonics in the frequency spectra of RRc and RRd
stars is very significant, as it is a characteristic signature of a
period doubling behaviour. Period doubling is a phenomenon which is
well known in many dynamical systems (Berg\'e et al. 1986). In the
context of stellar pulsations, it results in an alternating light
curve in which even and odd pulsation cycles have different shapes.
Such a light curve repeats itself after {\it two} pulsation periods,
not one. The appearance of subharmonics in the Fourier spectrum is a
direct consequence of this property (e.g. Smolec \& Moskalik 2012,
their Fig.\,3).

Light curves with alternating deep and shallow minima have been
known for decades in RV~Tauri stars. Only in the 1980s were they
interpreted as resulting from period doubling (Buchler \& Kov\'acs
1987; Kov\'acs \& Buchler 1988). More recently, the period doubling
effect has been discovered in two other types of pulsating variables
-- in Blazhko-modulated RRab stars (Kolenberg et al. 2010; Szab\'o
et al. 2010) and in BL~Herculis stars (Smolec et al. 2012). In the
former case, its strength varies significantly over the Blazhko
cycle, and does not repeat from one Blazhko cycle to the next. The
picture is different in the BL~Herculis stars, where the alternating
light curve is strictly periodic (repetitive).

As was first recognized by Moskalik \& Buchler (1990),  period
doubling in pulsating variables can be caused by a half-integer
resonance between the modes. Through hydrodynamical modelling, its
origin in RV~Tauri stars was traced back to the 5:2 resonance with
the second overtone (Moskalik \& Buchler 1990), in RRab stars to the
9:2 resonance with the 9th overtone (Koll\'ath et al. 2011) and in
BL~Herculis stars to the 3:2 resonance with the first overtone
(Buchler \& Moskalik 1992; Smolec et al. 2012). In these three
classes of pulsators the period doubling affects the dominant mode.
In the case of the {\it Kepler} RRc stars and the other RRc and RRd
stars of Table\,\ref{TRRcother}, however, we do not observe this
phenomenon in the main radial mode(s). Instead, it is the puzzling
{\it secondary nonradial mode} with a period ratio of $\sim\! 0.61$
to the first overtone, that shows a period doubling behaviour.
Therefore, as we do not know the identity of this mode, it is
unclear which resonance (if any) is responsible for the period
doubling in the RRc and RRd variables.

\subsection{Variability of the secondary mode}\label{Sdisvar}

In all four RRc stars observed with {\it Kepler}, the secondary
nonradial mode, $f_2$, and its subharmonics at $\sim\! 1/2f_2$ and
$\sim\! 3/2f_2$ display very strong variations of amplitudes and
phases. Variability of the $f_2$ mode has also been found in other
RRc stars observed from space (Szab\'o et al. 2014; Moln\'ar et al.
in preparation), indicating that this is yet another common property
of all RRc pulsators.

Thanks to the superb quality of the {\it Kepler} data, we have been
able to study this phenomenon with unprecedented detail. We have
shown that variability of the secondary mode is quasiperiodic and
occurs on a timescale of $10-\!200$\,d, depending on the star. In
the frequency domain, it causes the mode to split into a quintuplet
of equally spaced peaks. Each of these peaks is broadened, which
reflects an irregular character of the modulation. The main radial
mode, $f_1$, also varies in a quasiperiodic way, but this modulation
is extremely small. In the frequency domain, the mode splits into a
quintuplet as well. In every star the separation between the
quintuplet components of $f_1$ and of $f_2$ is the same, which
proves that both modes are variable with the same timescale.

The finding that the secondary mode and the main radial mode are
both modulated on a common timescale has implications for
understanding of the nature of the $f_2$ multiplet. Because $f_1$
corresponds to a radial mode ($\ell=0$), its quintuplet structure
cannot be caused by rotational splitting. It can be interpreted only
in one way -- as resulting from a true, physical modulation of the
mode's amplitude and phase. The same frequency splitting and thus
the same variability timescale of $f_1$ and $f_2$ indicates that
modulations of the two modes are not independent. This in turn
implies that quintuplet pattern of $f_2$ cannot correspond to a
multiplet of nonradial modes, either. Accepting such a picture would
make it difficult to understand why beating of rotationally split
multiplet ($f_2$) should occur on the same timescale as the true
modulation of another mode ($f_1$). We believe that such a
coincidence is highly unlikely. This leads us to the conclusion that
the observed modulation of the secondary mode, $f_2$, is due not to
beating, but just as in the case of $f_1$, to a true physical
modulation of a single pulsation mode.

The same modulation timescale of $f_1$ and $f_2$ implies that both
modes must be part of the same dynamical system, in other words that
they must interact with each other. We can only speculate what the
nature of this interaction might be. Cross-saturation is the most
obvious possibility that comes to mind. In physical terms, the two
modes compete for the same driving ($\kappa$-mechanism in the He$^+$
partial ionization zone) and when the amplitude of one mode
decreases, the amplitude of the other mode can increase. This kind
of coupling {\it has to occur} always when the two modes use the
same driving source. It also predicts that the amplitude variations
of the two modes should be anticorrelated. This is what we observe.

\section{Summary and conclusions}\label{Ssummary}

In this paper we present an in-depth analysis of four first overtone
RR~Lyrae stars (RRc stars) observed with the {\it Kepler} space
telescope: KIC\,4064484, KIC\,5520878, KIC\,8832417 and
KIC\,9453114. For our study we used the Long Cadence data (30\,min
sampling) gathered between Q0 and Q10, with a total timebase of 774
to 880\,d, depending on the star. Our most important findings can be
summarized as follows:

\begin{itemize}
\item None of the studied {\it Kepler} RRc stars displays a classical
      Blazhko effect, with a nearly coherent periodic (or
      multi-periodic) modulation of the amplitude and phase of the
      dominant radial mode.

\item In every {\it Kepler} RRc star we detect a secondary mode,
      $f_2$, with a period ratio of $\sim\! 0.61$ to the first
      radial overtone. The mode has a very low amplitude, more than
      20 times below that of the dominant radial mode.

\item Secondary modes with similar period ratios are also present in 19
      other RR~Lyrae variables. These stars are either of RRc or of
      RRd type, but never of RRab type. Apparently, the excitation
      of the secondary mode is somehow connected with the excitation
      of the first radial overtone. The observed period ratios are
      in a very narrow range of $0.602-0.632$, defining a new class
      of multimode pulsators. Including the four RRc stars studied
      in this paper, this class has currently 23 members.

\item The period ratio of $\sim\! 0.61$ is also observed in Classical
      Cepheids of the Magellanic Clouds. The stars are either single
      mode first overtone pulsators or F+1O double-mode pulsators,
      but never single mode fundamental pulsators. This property is
      the same in the case of Cepheids and in the case of RR~Lyrae stars.

\item In neither RR~Lyrae stars nor Cepheids can the period ratio of
      $\sim\! 0.61$ be reproduced by two radial modes. In both
      types of variables the secondary mode must be nonradial.

\item In every {\it Kepler} RRc star we detect at least one subharmonic
      of the secondary mode, at $\sim\! 1/2f_2$ or at $\sim\!
      3/2f_2$. Similar subharmonics have also been found recently in
      several other RRc and RRd variables observed from space
      (Moln\'ar et al. in preparation). They have also been
      retrospectively identified in two more RRd stars observed from
      space. Detection of subharmonics of $f_2$ is a signature of
      period doubling of this mode. After RV~Tauri, Blazhko RRab and
      BL~Herculis stars, the RRc and RRd stars are now the fourth
      group of pulsators in which period doubling has been found.
      Contrary to the former three types of variables, in the RRc
      and RRd stars the period doubling affects not the primary, but
      the secondary mode of pulsation.

\item Judging from the results of space photometric observations,
      the excitation of the secondary mode with the period ratio of
      $\sim\! 0.61$ to the first overtone and concomitant period
      doubling of this mode must be a common phenomenon in RRc
      and RRd variables.

\item In every {\it Kepler} RRc star the amplitude and phase of the
      secondary mode and its subharmonics are strongly variable,
      with timescales of $10-200$\,d. The main radial mode varies
      on the same timescale, but with an extremely low amplitude.
      Its variability can be detected only with high-precision
      photometry collected from space.

\item In three {\it Kepler} RRc stars even more periodicities are
      detected, all with amplitudes well below 1\,mmag. One of the
      low amplitude modes discovered in KIC\,5520878 can be
      identified with the radial second overtone, but all others
      must be nonradial. Many of these modes appear at frequencies
      that are below that of the radial fundamental mode. As such,
      they cannot be acoustic oscillations ($p$-modes), but must be
      classified as gravity modes ($g$-modes). This is the first
      detection of such modes in the RR~Lyrae stars. The result is
      very surprising. From the theoretical point of view, the
      $g$-modes are not expected to be excited in the RR~Lyrae
      variables, because they are all strongly damped in the
      radiative interior of the star.
\end{itemize}

At the time of writing, several additional RR Lyrae stars have been
found in the {\it Kepler} field, thanks to the efforts of many
dedicated individuals, and often as a by-product of studies devoted
to other objects.  For example, short period eclipsing binaries can
have periods and amplitudes in the range of RR~Lyrae stars (e.g.,
Rappaport et al. 2013; Szab\'o et al. in~preparation).  Also, the
citizen science project PlanetHunters (www.planethunters.org) has
yielded some additional RR Lyrae stars in the {\it Kepler} field.
Among the new finds are several RRc stars.

The sample of space-observed RRc and RRd variables will be further
expanded by the {\it Kepler} K2 mission, which is already underway
(Howell et al. 2014). It is observing in the plane of the ecliptic,
and will switch the field of view every 3\,months. With this
strategy, a variety of stellar populations will be sampled,
including not only the Galactic halo and the thick disc but also
several globular clusters and the Galactic Bulge. The K2 mission can
observe hundreds of RR~Lyrae variables, among them tens of RRc and
of RRd type (Moln\'ar, Plachy \& Szab\'o 2014). The mission will
increase greatly the number such objects studied from space. It will
be interesting to see whether they show characteristics similar to
the four RRc variables discussed in this paper.

When writing of this manuscript was almost completed, a new paper on
RRc stars has been submitted by Netzel et al. (2014). The authors
have analyzed the OGLE-III photometry of the Galactic Bulge. In
$\sim 3$ per cent of the studied variables, they have detected a
low-amplitude secondary mode with the period ratio of $\sim\! 0.61$
to the first radial overtone. This result confirms our conclusion
that excitation of this puzzling mode is a common phenomenon in RRc
stars.

\section*{Acknowledgements}

Funding for this Discovery mission was provided by NASA's Science
Mission Directorate. The authors gratefully acknowledge the entire
{\it Kepler} team, whose outstanding efforts have made these results
possible. This research has been supported by the Polish NCN through
grant no. DEC-2012/05/B/ST9/03932. It has also been supported by the
"Lend\"ulet-2009 Young Researchers" Program of the Hungarian Academy
of Sciences, the Hungarian OTKA grant K83790, the National Science
Foundation grant no. NSF PHY05-51164, the European Community's
Seventh Framework Programme (FP7/2007-2013) grants no. 269194
(IRSES/ASK), 312844 (SPACEINN) and 338251 (StellarAges), the ESA
PECS Contract no. 4000110889/14/NL/NDe, the Ministry of Science and
Technology (Taiwan) grant no. MOST101-2112-M-008-017-MY3, the Korea
Astronomy and Space Science Institute Project no. 2014-1-400-06,
supervised by the Ministry of Science, ICT and Future Planning and
by the Polish NCN grant no. 2011/03/B/ST9/02667. KK is currently a
Marie Curie Fellow, grateful for the support of grant
PIOF-GA-2009-255267 (SAS-RRL). RSz acknowledges the University of
Sydney IRCA grant. JMN wishes to thank the Camosun College Faculty
Association for financial assistance.

\appendix
\section{Time-dependent prewhitening}\label{Sappendix1}

\begin{figure}
\vskip 2.95truecm
\centering
\resizebox{\hsize}{!}{\includegraphics{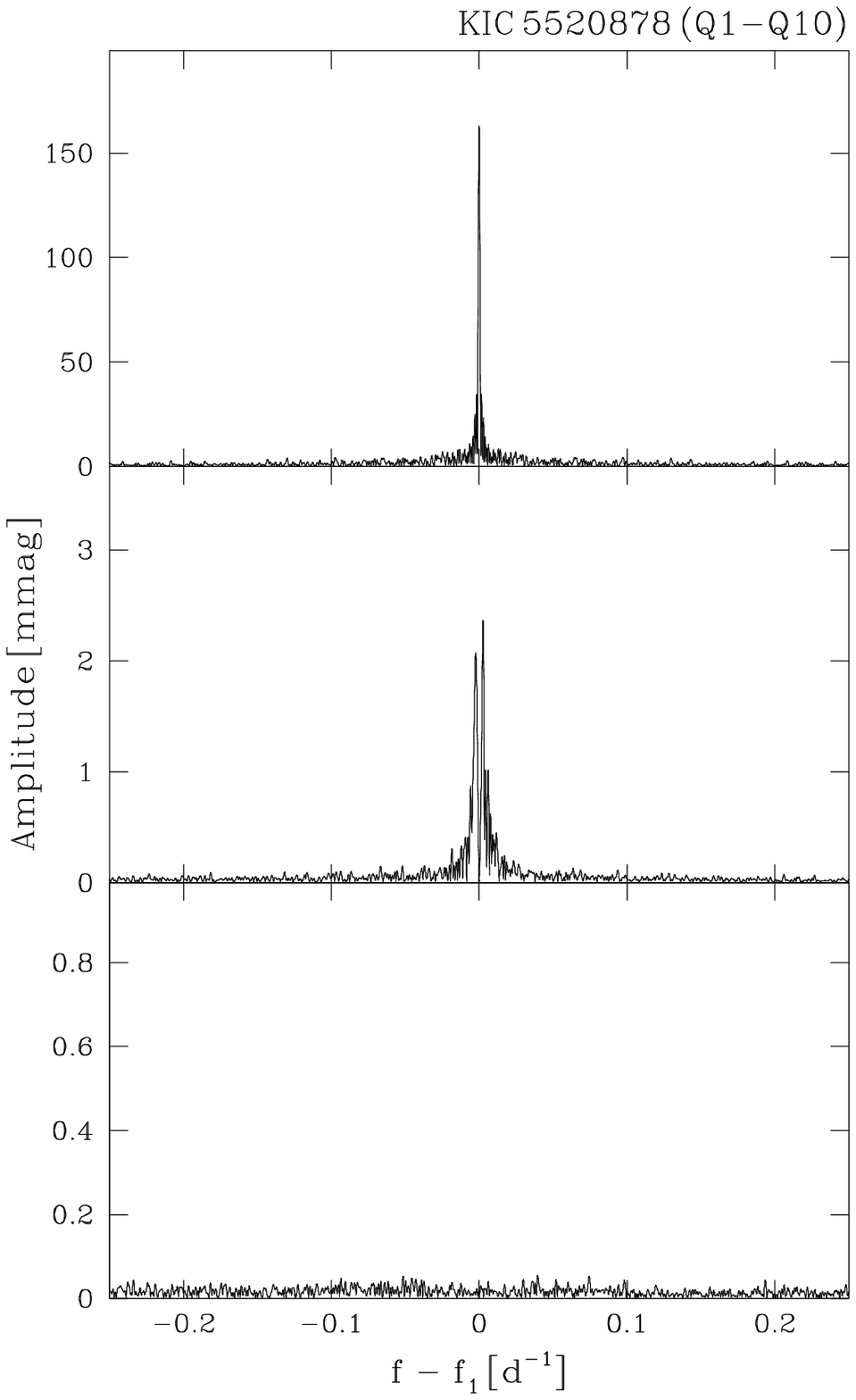}}
\caption{Prewhitening of the main frequency in KIC\,5520878. Upper
         panel: Fourier transform of the original $K\!p$ magnitude
         light curve (quarters Q1 -- Q10). Middle panel: FT after
         standard prewhitening. Bottom panel: FT after
         time-dependent prewhitening ($\Delta t = 10$\,d).}
\label{FA1}
\end{figure}

In the standard prewhitening procedure a periodic signal is removed
from the light curve by subtracting a sine function with constant
frequency ($f_1$), amplitude ($A_1$) and phase ($\phi_1$). The
values of $f_1$, $A_1$ and $\phi_1$ are determined from the data
with a least-squares fit. The method works very well when the
signal's frequency, amplitude and phase are indeed non-variable.
However, when this condition is violated, as is the case for the
{\it Kepler} RRc stars, the standard procedure fails and leaves
residual power in the frequency spectrum of the prewhitened data.
This is illustrated in Fig.\,\ref{FA1}, where we display the
prewhitening sequence for the dominant mode of KIC\,5520878. The
residuals in the FT (middle panel) have an amplitude of 2.4\,mmag.
This is only $\sim$1.5~per~cent of the original peak's amplitude,
but in context of {\it Kepler} photometry this is huge.

To remedy this situation, we have developed a novel method that we
call the {\it time-dependent prewhitening}. In this procedure, we
subtract from the light curve a sine function not with constant, but
with {\it varying} amplitude and phase. To model this variability,
we adopt $A_1(t)$ and $\phi_1(t)$ determined from the data
(Fig.\,\ref{FRRcA1}) by the time-dependent Fourier analysis
(Kov\'acs et al. 1987). The frequency of the mode, $f_1$, is kept
fixed. In the bottom panel of Fig.\,\ref{FA1} we display the results
of the time-dependent prewhitening for KIC\,5520878. The dominant
frequency is now removed entirely, down to the noise level of the FT
(55\,$\mu$mag).

\begin{figure*}
\vskip 3.05truecm
\centering
\resizebox{0.75\hsize}{!}{\includegraphics{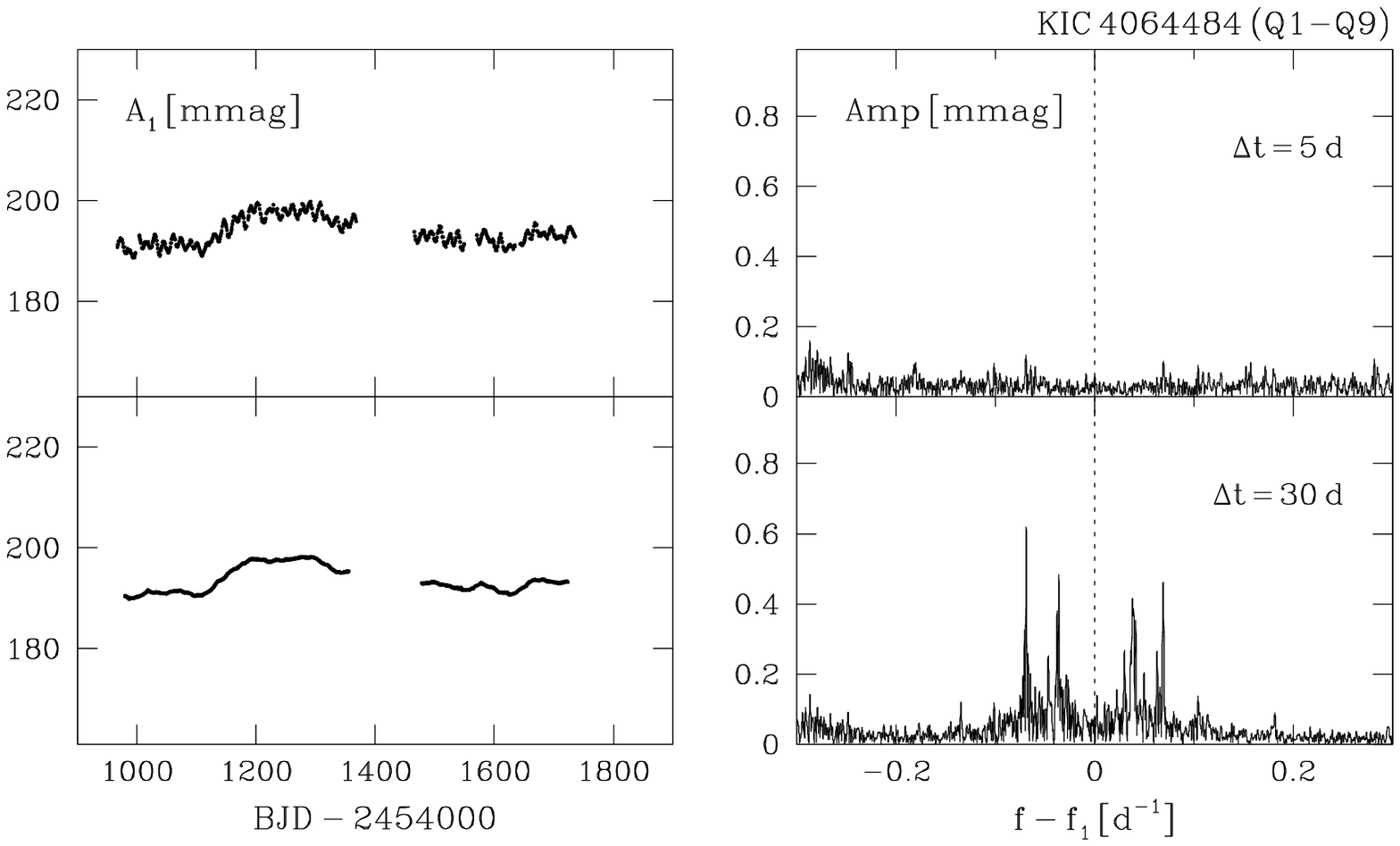}}
\caption{Time-dependent prewhitening of the dominant mode of
         KIC\,4064484 for the light curve segment length of $\Delta
         t\! =\! 5$\,d and $\Delta t\! =\! 30$\,d. Left column:
         variations of the amplitude of the mode, $A_1(t)$,
         determined with the time-dependent Fourier analysis. Right
         column: FT of the prewhitend light curve. Frequency of the
         removed dominant mode, $f_1$, is indicated by the dashed
         line.}
\label{FA2}
\end{figure*}

Time-dependent Fourier analysis and, consequently, also
time-dependent prewhitening, have one free parameter: the length of
the light curve segment, $\Delta t$. The choice of this parameter
determines the time resolution of the method. Variations occuring on
timescales shorter than $\Delta t$ will not be captured by the
time-dependent Fourier analysis and consequently, they will not be
subtracted by the prewhitening procedure.

We illustrate this property in Fig.\,\ref{FA2}, where we present
prewhitening of the dominant frequency of KIC\,4064484. In the upper
panel we plot the results for $\Delta t = 5$\,d. The time-dependent
Fourier analysis (left column) captures both the long-term trend and
the rapid quasi-periodic modulation of the amplitude. Using this
$A_1(t)$ and concomitant $\phi_1(t)$ (not shown), the time-dependent
prewhitening removes from the FT all power associated with the mode
(right column). The results for $\Delta t\! =\! 30$\,d are quite
different. The time-dependent Fourier analysis now captures only the
long-term trend, but the rapid modulation is averaged out. After the
time-dependent prewhitening, we find four residual peaks in the FT.
They are placed symmetrically around the (removed) central
frequency, $f_1$, forming an equidistant frequency multiplet. These
peaks are the Fourier representation of the quasi-periodic
modulation of the mode, which for $\Delta t = 30$\,d is not captured
by the method and not removed during prewhitening.

In the language of the Fourier analysis, $\Delta t = 5$\,d
corresponds to a bandwidth of 0.2\,d$^{-1}$ ($1/\Delta t$). This is
broader than the frequency multiplet in Fig.\,\ref{FA2}. As a
result, the signal reconstructed with the time-dependent Fourier
analysis captures {\it all} the Fourier power associated with the
mode and {\it all} of it can be removed. For $\Delta t = 30$\,d, the
bandwidth is only 0.0333\,d$^{-1}$. This is narrow enough to isolate
only the central peak of the multiplet. Consequently, only this peak
is subtracted in the prewhitening process.


\begin{thebibliography}{}
\bibitem[\protect\citeauthoryear{Bailey}{1902}]{Bailey}
         Bailey S.\,I., 1902, Harv. Coll. Observ. Annals, 38,~1                                 
\bibitem[\protect\citeauthoryear{Benk\H{o} et al.}{2010}]{Benko10}
         Benk\H{o} J.\,M., Kolenberg K., Szab\'o R. et al., 2010, MNRAS, 409,~1585              
\bibitem[\protect\citeauthoryear{Benk\H{o}, Szab\'o \& Papar\'o}{Benk\H{o} et al.}{2011}]
         {Benko11} Benk\H{o} J.\,M., Szab\'o R., Papar\'o M., 2011, MNRAS, 417,~974             
\bibitem[\protect\citeauthoryear{Benk\H{o} et al.}{2014}]{Benko14}
         Benk\H{o} J.\,M., Plachy E., Szab\'o R., Moln\'ar L., Koll\'ath Z., 2014,
         ApJS, 213,~31                                                                          
\bibitem[\protect\citeauthoryear{Berdnikov et al.}{1997}]{Berd97}
         Berdnikov L.\,N., Ignatova V.\,V., Pastukhova E.\,N., Turner~D.\,G., 1997,
         Astron. Lett., 23,~177                                                                 
\bibitem[\protect\citeauthoryear{Berg\'e, Pomeau \& Vidal}{Berg\'e et al.}{1986}]{BPV86}
         Berg\'e P., Pomeau Y., Vidal Ch., 1986, Order within Chaos (New York: Wiley)           
\bibitem[\protect\citeauthoryear{Blazhko}{1907}]{Blazhko}
         Blazhko S., 1907, Astron. Nachr., 175,~325                                             
\bibitem[\protect\citeauthoryear{Borucki et al.}{2010}]{Bor10}
         Borucki W.\,J., Koch D.\,G., Basri G. et al., 2010, Sci, 327,~977                      
\bibitem[\protect\citeauthoryear{Bowman \& Kurtz}{2014}]{BK14}
         Bowman D.\,M., Kurtz D.\,W., 2014, MNRAS, 444,~1909                                    
\bibitem[\protect\citeauthoryear{Brown et al.}{2011}]{Brown11}
         Brown T.\,M., Latham D.\,W. et al., 2012, AJ, 142,~112                                 
\bibitem[\protect\citeauthoryear{Buchler \& Kov\'acs}{1987}]{BK87}
         Buchler J.\,R., Kov\'acs G., 1987, ApJL, 320,~L57                                      
\bibitem[\protect\citeauthoryear{Buchler \& Moskalik}{1992}]{BM92}
         Buchler J.\,R., Moskalik P., 1992, ApJ, 391,~736                                       
\bibitem[\protect\citeauthoryear{Caldwell et al.}{2010}]{Cald10}
         Caldwell D.\,A., Kolodziejczak J.\,J., Van Cleve J.\,E. et al., 2010, ApJ, 713,~L92    
\bibitem[\protect\citeauthoryear{Chadid}{2012}]{Chad12}
         Chadid M., 2012, A\&A, 540,~A68                                                        
\bibitem[\protect\citeauthoryear{Derekas et al.}{2004}]{Der04}
         Derekas A., Kiss L.\,L., Udalski A., Bedding T.\,R., Szatm\'ary~K., 2004,
         MNRAS, 354,~821                                                                        
\bibitem[\protect\citeauthoryear{Dziembowski}{2012}]{WD12}
         Dziembowski W.\,A., 2012, Acta Astron., 62,~323                                        
\bibitem[\protect\citeauthoryear{Dziembowski \& Cassisi}{1999}]{DC99}
         Dziembowski W.\,A., Cassisi S., 1999, Acta Astron., 49,~371                            
\bibitem[\protect\citeauthoryear{Dziembowski \& Smolec}{2009}]{DS09}
         Dziembowski W.\,A., Smolec R., 2009, in Guzik J.\,A., Pradley P.\,A., eds,
         AIP Conf. Proc., 1170,~83                                                              
\bibitem[\protect\citeauthoryear{Gilliland et al.}{2010}]{Gill10}
         Gilliland R.\,L., Jenkins J.\,M., Borucki W.\,J. et al., 2010, ApJ, 713,~L160          
\bibitem[\protect\citeauthoryear{Gruberbauer et al.}{2007}]{Grub07}
         Gruberbauer M., Kolenberg K., Rowe J.\,F. et al., 2007, MNRAS, 379,~1498               
\bibitem[\protect\citeauthoryear{Guggenberger et al.}{2012}]{Gugg12}
         Guggenberger E., Kolenberg K., Nemec J.\,M. et al., 2012, MNRAS, 424,~649              
\bibitem[\protect\citeauthoryear{Haas et al.}{2010}]{Haas10}
         Haas M.\,R., Batalha N.\,M., Bryson S.\,T. et al., 2010, ApJ, 713,~L115                
\bibitem[\protect\citeauthoryear{Howell et al.}{2014}]{HSH14}
         Howell S.\,B., Sobeck Ch., Haas M. et al., 2014, PASP, 126,~398                        
\bibitem[\protect\citeauthoryear{Jenkins et al.}{2010a}]{Jen10a}
         Jenkins J.\,M., Caldwell D.\,A., Chandrasekaran H. et al., 2010a, ApJ, 713,~L87        
\bibitem[\protect\citeauthoryear{Jenkins et al.}{2010b}]{Jen10b}
         Jenkins J.\,M., Caldwell D.\,A., Chandrasekaran H. et al., 2010b, ApJ, 713,~L120       
\bibitem[\protect\citeauthoryear{Jurcsik et al.}{2001}]{Jur01}
         Jurcsik J., Clement C., Geyer E.\,H., Domsa I., 2001, AJ, 121,~951                     
\bibitem[\protect\citeauthoryear{Jurcsik et al.}{2009}]{Jur09}
         Jurcsik J., S\'odor \'A., Szeidl B. et al., 2009, MNRAS, 400,~1006                     
\bibitem[\protect\citeauthoryear{Jurcsik et al.}{2012}]{Jur12}
         Jurcsik J., Hajdu G., Szeidl B. et al., 2012, MNRAS, 419,~2173                         
\bibitem[\protect\citeauthoryear{Koch et al.}{2010}]{Koch10}
         Koch D.\,G., Borucki W.\,J., Basri G. et al., 2010, ApJ, 713,~L79                      
\bibitem[\protect\citeauthoryear{Kolenberg et al.}{2010}]{Kolen10}
         Kolenberg K., Szab\'o R., Kurtz D.\,W. et al., 2010, ApJL, 713,~198                    
\bibitem[\protect\citeauthoryear{Kolenberg et al.}{2011}]{Kolen11}
         Kolenberg K., Bryson S., Szab\'o R. et al., 2011, MNRAS, 411,~878                      
\bibitem[\protect\citeauthoryear{Kolenberg et al.}{2014}]{Kol14}
         Kolenberg K., Kurucz R.\,L., Stellingwerf R. et al., 2014, in Guzik J.\,A.,
         Chaplin W.\,J., Handler~G., Pigulski A., eds, IAU~Symp.~301, p.~257                    
\bibitem[\protect\citeauthoryear{Koll\'ath et al.}{2002}]{KBSC02}
         Koll\'ath Z., Buchler J.\,R., Szab\'o R., Csurby Z., 2002, A\&A, 385,~932              
\bibitem[\protect\citeauthoryear{Koll\'ath, Moln\'ar \& Szab\'o}{Koll\'ath et al.}{2011}]{KMS11}
         Koll\'ath Z., Moln\'ar L., Szab\'o R., 2011, MNRAS, 414,~1111                          
\bibitem[\protect\citeauthoryear{Kov\'acs}{2001}]{Kov00}
         Kov\'acs G. , 2001, in Takeuti M., Sasselov D.\,D., eds, Stellar Pulsation --
         Nonlinear Studies, Astrophys. Space Sci. Library, Vol.~257 (Dordrecht: Kluwer),
         p.~61                                                                                  
\bibitem[\protect\citeauthoryear{Kov\'acs \& Buchler}{1988}]{KB88}
         Kov\'acs G., Buchler J.\,R., 1988, ApJ, 334,~971                                       
\bibitem[\protect\citeauthoryear{Kov\'acs, Buchler \& Davis}{Kov\'acs et al.}{1987}]{KBD87}
         Kov\'acs G., Buchler J.\,R., Davis C.\,G., 1987, ApJ, 319,~247                         
\bibitem[\protect\citeauthoryear{Le\,Borgne et al.}{2007}]{LB07}
         Le\,Borgne J.\,F., Paschke A., Vandenbroere J. et al., 2007, A\&A, 476,~307            
\bibitem[\protect\citeauthoryear{Lub}{1977}]{Lub77}
         Lub J., 1977, A\&AS, 29,~345                                                           
\bibitem[\protect\citeauthoryear{Mizerski}{2003}]{Mizer03}
         Mizerski T., 2003, Acta Astron., 53,~307                                               
\bibitem[\protect\citeauthoryear{Moln\'ar et al.}{2012}]{Moln12}
         Moln\'ar L., Koll\'ath Z., Szab\'o R. et al., 2012, ApJL, 757,~L13
\bibitem[\protect\citeauthoryear{Moln\'ar, Plachy \& Szab\'o}{Moln\'ar et al.}{2014}]
         {Molnar14} Moln\'ar L., Plachy E., Szab\'o R., 2014, IBVS, 6108                        
\bibitem[\protect\citeauthoryear{Morgan, Simet \& Bargenquast}{Morgan et al.}{1998}]{MSB98}
         Morgan S.\,M., Simet M., Bargenquast S., 1998, Acta Astron., 48,~341                   
\bibitem[\protect\citeauthoryear{Moskalik}{2013}]{Mos13}
         Moskalik P., 2013, in Su\'arez J.\,C., Garrido R., Balona L.\,A.,
         Christensen-Dalsgaard J., eds, Astrophysics and Space Science Proc.~31, p.~103         
\bibitem[\protect\citeauthoryear{Moskalik}{2014}]{Mos14}
         Moskalik P., 2014, in Guzik J.\,A., Chaplin W.\,J.,
         Handler~G., Pigulski A., eds, IAU~Symp.~301, p.~249                                    
\bibitem[\protect\citeauthoryear{Moskalik \& Buchler}{1990}]{MB90}
         Moskalik P., Buchler J.\,R., 1990, ApJ, 355,~590                                       
\bibitem[\protect\citeauthoryear{Moskalik \& Ko{\l}aczkowski}{2008}]{MK08}
         Moskalik P., Ko{\l}aczkowski Z., 2008, Comm. in Asteroseismology, 157,~343             
\bibitem[\protect\citeauthoryear{Moskalik \& Ko{\l}aczkowski}{2009}]{MK09}
         Moskalik P., Ko{\l}aczkowski Z., 2009, MNRAS, 394,~1649                                
\bibitem[\protect\citeauthoryear{Moskalik et al.}{2013}]{MSK13}
         Moskalik P., Smolec R., Kolenberg K. et al., 2013, in Su\'arez J.\,C.,
         Garrido R., Balona L.\,A., Christensen-Dalsgaard J., eds,
         Astrophysics and Space Science Proc.~31, poster No~34 (arXiv:\,1208.4251)              
\bibitem[\protect\citeauthoryear{Nagy \& Kov\'acs}{2006}]{NK06}
         Nagy A., Kov\'acs G., 2006, A\&A, 454,~257
\bibitem[\protect\citeauthoryear{Nemec et al.}{2011}]{Nem11}
         Nemec J.\,M., Smolec R., Benk{\H o} J.\,M. et al., 2011, MNRAS, 417,~1022              
\bibitem[\protect\citeauthoryear{Nemec et al.}{2013}]{Nem13}
         Nemec J.\,M., Cohen J.\,G., Ripepi V. et al., 2013, ApJ, 773,~181                      
\bibitem[\protect\citeauthoryear{Netzel, Smolec \& Moskalik}{Netzel et al.}{2014}]{NSM14}
         Netzel H., Smolec R., Moskalik P., 2014, MNRAS, in press, arXiv:1411.3155              
\bibitem[\protect\citeauthoryear{Olech \& Moskalik}{2009}]{OM09}
         Olech A., Moskalik P., 2009, A\&A, 494,~L17                                            
\bibitem[\protect\citeauthoryear{Olech et al.}{2001}]{Olech11}
         Olech A., Kaluzny J., Thompson I.\,B. et al., 2001, MNRAS, 321,~421                    
\bibitem[\protect\citeauthoryear{Payne-Gaposchkin \& Gaposchkin}{1966}]{PGG66}
         Payne-Gaposchkin C., Gaposchkin S., 1966, Smithsonian Contrib. Astrophys., 9~1         
\bibitem[\protect\citeauthoryear{Pigulski}{2014}]{Pig14}
         Pigulski A., 2014, in Guzik J.\,A., Chaplin W.\,J., Handler~G., Pigulski A., eds,
         IAU~Symp.~301, p.~31                                                                   
\bibitem[\protect\citeauthoryear{Popielski, Dziembowski \& Cassisi}{Popielski et al.}{2000}]
         {Popiel00} Popielski B.\,L., Dziembowski W.\,A.,  Cassisi S., 2010,
         Acta Astron., 50,~491                                                                  
\bibitem[\protect\citeauthoryear{Rappaport et al.}{2013}]{Rap13}
         Rappaport S., Deck K., Levine A. et al., 2013, ApJ, 768,~33                            
\bibitem[\protect\citeauthoryear{Simon \& Lee}{1981}]{SL81}
         Simon N., Lee A.\,S., 1981, ApJ, 248,~291                                              
\bibitem[\protect\citeauthoryear{Smolec \& Moskalik}{2008a}]{SM08a}
         Smolec R., Moskalik P., 2008a, Acta Astron., 58,~193                                   
\bibitem[\protect\citeauthoryear{Smolec \& Moskalik}{2008b}]{SM08b}
         Smolec R., Moskalik P., 2008b, Acta Astron., 58,~233                                   
\bibitem[\protect\citeauthoryear{Smolec \& Moskalik}{2010}]{SM10}
         Smolec R., Moskalik P., 2010, A\&A, 524,~A40                                           
\bibitem[\protect\citeauthoryear{Smolec \& Moskalik}{2012}]{SM12}
         Smolec R., Moskalik P., 2012, MNRAS, 426,~108                                          
\bibitem[\protect\citeauthoryear{Smolec et al.}{2012}]{SSM12}
         Smolec R., Soszy\'nski I., Moskalik P. et al., 2012, MNRAS, 419,~2407                  
\bibitem[\protect\citeauthoryear{Soszy\'nski et al.}{2008}]{Sosz08}
         Soszy\'nski I., Poleski R., Udalski A. et al., 2008, Acta Astron., 58,~163             
\bibitem[\protect\citeauthoryear{Soszy\'nski et al.}{2009}]{Sosz09}
         Soszy\'nski I., Udalski A., Szyma\'nski M.\,K. et al., 2009, Acta Astron., 59,~1       
\bibitem[\protect\citeauthoryear{Soszy\'nski et al.}{2010}]{Sosz10}
         Soszy\'nski I., Poleski R., Udalski A. et al., 2010, Acta Astron., 60,~17              
\bibitem[\protect\citeauthoryear{Soszy\'nski et al.}{2011a}]{Sosz11a}
         Soszy\'nski I., Dziembowski W.\,A., Udalski A. et al., 2011a, Acta Astron., 61,~1      
\bibitem[\protect\citeauthoryear{Soszy\'nski et al.}{2011b}]{Sosz11b}
         Soszy\'nski I., Udalski A., Pietrukowicz P. et al., 2011b, Acta Astron., 61,~285       
\bibitem[\protect\citeauthoryear{S\"uveges et al.}{2012}]{Suv12}
         S\"uveges M., Sesar B., V\'aradi M. et al., 2012, MNRAS, 424,~2528                     
\bibitem[\protect\citeauthoryear{Szab\'o}{2014}]{Szabo14}
         Szab\'o R., 2014, in Guzik J.\,A., Chaplin W.\,J., Handler~G., Pigulski A., eds,
         IAU~Symp.~301, p.~241                                                                  
\bibitem[\protect\citeauthoryear{Szab\'o et al.}{2010}]{SKM10}
         Szab\'o R., Koll\'ath Z., Moln\'ar L. et al., 2010, MNRAS, 409,~1244                   
\bibitem[\protect\citeauthoryear{Szab\'o et al.}{2014}]{SBP14}
         Szab\'o R., Benk{\H o} J.\,M., Papar\'o M. et al., 2014, A\&A, 570,~A100               
\bibitem[\protect\citeauthoryear{Tsesevich}{1975}]{Tses75}
         Tsesevich V.\,P., 1975, in Kukarkin B.\,V., ed, Pulsating Stars, IPST Astrophys.
         Library (New York: Willey), p.~144                                                     
\bibitem[\protect\citeauthoryear{Van~Hoolst, Dziembowski \& Kawaler}{Van Hoolst et al.}
         {1998}]{VHDK98} Van~Hoolst T., Dziembowski W.\,A., Kawaler S.\,D., 1998,
         MNRAS, 297,~536                                                                        
\bibitem[\protect\citeauthoryear{Walker}{1994}]{Wal94}
         Walker A.\,R., 1994, AJ, 108,~555                                                      
\end{thebibliography}
\end{document}